\documentclass[12pt,aip]{article}
\usepackage{graphicx,bbold,epsfig,amsmath,xcolor,ulem,nccmath,lipsum,tikz,array,lineno,amssymb,comment,microtype}
\microtypesetup{protrusion=false}
\usepackage[nodisplayskipstretch]{setspace}
\usepackage{fancyhdr}
\usepackage[useregional]{datetime2}

\usepackage[
backend=biber,                
maxnames=6,                         
bibstyle=phys,
citestyle=numeric-comp,
hyperref=true,                  
sortcites=true,
sorting=none,maxbibnames=99
]{biblatex}
\bibliography{references}

\begin{document}

\vskip 2 cm

%\maketitle

{\color{black}\huge \bf Machine Learning Link Inference of Noisy Delay-coupled Networks with Opto-Electronic Experimental Tests}
%\maketitle

\begin{center}
{\large
   Amitava Banerjee$^{\dagger 1,2}$, Joseph D. Hart$^{\ast 3}$, Rajarshi Roy$^{1,2,4}$, Edward Ott$^{1,2,5}$ }
\end{center}

\vskip 2cm
1. Department of Physics, University of Maryland, College Park, Maryland 20742, U.S.A.\\
2. Institute for Research in Electronics and Applied Physics, University of Maryland, College Park, Maryland 20742, U.S.A.\\
3. Optical Sciences Division, US Naval Research Laboratory, Washington, DC 20375, U.S.A.\\
4. Institute for Physical Science and Technology, University of Maryland, College Park, Maryland 20742, U.S.A. \\
5. Department of Electrical and Computer Engineering, University of Maryland, College Park, Maryland 20742, U.S.A.

\vskip 2 cm

{$^{\dagger}$ Corresponding author to contact for codes, data, and accessible versions of the presented materials, emails: amitava8196@gmail.com, amitavab@umd.edu, pronouns; he/him, they/them}

{$^{\ast}$ Corresponding author to contact for details of the experiments performed, email: joseph.hart@nrl.navy.mil}

%{DISTRIBUTION A. Approved for public release, distribution is unlimited.}

\renewcommand{\baselinestretch}{2}
\small\normalsize
\newpage
%\linenumbers
\begin{abstract}
\hskip -.2in
\noindent

%1. Both referees 2 and 3 expressed concerns regarding our use of the illustrative assumption that the number of links L is known. As described in detail in our response to referee 2, we now address this concern of referees 2 and 3 in revised sections 2.3 (last two paragraphs), 4.1 (last paragraph), and newly-added figure 8.
%2. Add after the newly-added section 4.3 "and newly-added figure 10"
%3. Add "in the revised section 2.3"

%5. Take it out

We devise a machine learning technique to solve the general problem of inferring network links that have time delays using only time series data of the network nodal states. This task has applications in many fields, e.g., from applied physics, data science, and engineering to neuroscience and biology. Our approach is to first train a type of machine learning system known as reservoir computing to mimic the dynamics of the unknown network. We then use the trained parameters of the reservoir system output layer to deduce an estimate of the unknown network structure. Our technique,  by its nature, is non-invasive, but is motivated by the widely-used invasive network inference method whereby the responses to active perturbations applied to the network are observed and employed to infer network links (e.g., knocking down genes to infer gene regulatory networks). We test this technique on experimental and simulated data from delay-coupled opto-electronic oscillator networks, {\color{black} with both identical and heterogeneous delays along the links}. We show that the technique often yields very good results, particularly if the system does not exhibit synchrony. We also find that the presence of dynamical noise can strikingly enhance the accuracy and ability of our technique, especially in networks that exhibit synchrony.

\end{abstract}

\newpage
\tableofcontents

\newpage

\section{Introduction}

Dynamically evolving complex networks are ubiquitous in natural and technological systems \cite{albert2002statistical}. Examples include food webs \cite{dunne2002food}, biochemical \cite{price2007biochemical,vidal2011interactome} and gene interaction \cite{baryshnikova2013genetic, costanzo2016global} networks, neural networks \cite{bassett2017network}, human interaction networks \cite{pastor2015epidemic}, and the internet, to mention a few. Inference of the structure of such networks from observation of their dynamics is a key issue with applications such as determination of the connectivity in nervous systems \cite{de2018connectivity,lynall2010functional,chavez2010functional}, mapping of interactions between genes \cite{sima2009inference} and proteins in biochemical networks \cite{albert2007network}, distinguishing relationships between species in ecological networks \cite{sander2017ecological,milns2010revealing}, understanding the causal dependencies between elements of the global climate \cite{runge2015identifying},  and charting of the invisible dark web of the internet \cite{venegas2018tracing}. In many of these problems, we can only passively observe time series data for the states of the network nodes and cannot actively perturb the systems in any way. Network inference for these cases has led to several different computational and statistical approaches, including Granger causality \cite{doi:https://doi.org/10.1002/9783527609970.ch17,zhou2013causal}, transfer entropy \cite{schreiber2000measuring}, causation entropy \cite{sun2015causal}, event timing analysis \cite{casadiego2018inferring}, Bayesian techniques \cite{sima2009inference,milns2010revealing,zou2009granger,peixoto2019network}, inversion of response functions \cite{ren2010noise, dahmen2016correlated}, random forest methods \cite{leng2019reconstructing}, and feature ranking methods \cite{leguia2019reconstructing}, among others.

In this work, we are interested in the common situation of dynamics that evolves through interactions mediated by the network links along which information transfer is subject to time delay {\color{black} and dynamical noise}. We propose and test, both experimentally and computationally, a machine learning methodology to infer these time-delayed network interactions. In doing this, we use only the sampled time series data of the network nodal states. We find that our method is successful in both experimental and computational tests, {\color{black}for a wide variety of network topologies, and for networks with either identical or heterogeneous delay times along their links}, provided the time series we use contains sufficient information for the networks to be inferred.

Applications of machine learning techniques for network inference have recently begun to be explored \cite{leng2019reconstructing,banerjee2019using,chen2019generative,cao2019network,frolov2019feed}. {\color{black} However, all of these treat networks without link delay. For example, Ref. \cite{frolov2019feed} uses machine learning, but considers inference of generalized synchronization, rather than inference of links, between two systems. In particular, to our knowledge, there is no paper in which the common general situation considered in our paper is treated by machine learning, namely, link inference in delay-coupled networks with arbitrary topology and noise. Furthermore, a key feature of our work is that, in contrast with Refs. \cite{leng2019reconstructing,banerjee2019using,chen2019generative,cao2019network,frolov2019feed}, we present an experimental validation with {\it known ground truth}.} Based on the surprising success of machine learning across a wide variety of data-based tasks \cite{goodfellow2016deep,carleo2019machine}, we believe machine learning is a particularly promising approach to the general network inference problem that we address.

Our approach is based on the demonstrated ability of machine learning for the prediction and analysis of dynamical time series data. In particular, we shall use a specific neural network architecture called Reservoir Computing (RC) \cite{tanaka2019recent,jaeger2004harnessing}, which has previously been used to analyze time series data from complex and chaotic systems for such tasks as forecasting spatiotemporally chaotic evolution \cite{pathak2018model,pathak2018hybrid,neofotistos2019machine,zimmermann2018observing}, determination of Lyapunov exponents and replication of chaotic attractors \cite{pathak2017using}, chaotic source separation \cite{krishnagopal2020separation,lu2020supervised}, and inference of networks (without link time-delays) \cite{banerjee2019using}. Reservoir computers have been implemented in a variety of platforms \cite{tanaka2019recent,chembo2020machine,markovic2020physics}, e.g., in photonic \cite{larger2017high,vinckier2015high,duport2016fully}, electronic and opto-electronic \cite{schrauwen2008compact,appeltant2011information} systems. In our technique, we first use time series data from an unknown delay-coupled network to train an RC to predict the future evolution of the network. We then employ this trained network to predict how the effect of imagined applied perturbations would spread through the network, thus enabling us to deduce the network structure. This approach allows us to retain the non-invasive nature of computational tools like the transfer entropy, while also retaining the conceptual advantage of invasive methods \cite{molinelli2013perturbation,das2020systematic,olsen2014inference}. 

%In a sense, through our methodology, we are able to invert the structure-dynamics relationship of complex networks \cite{nishikawa2017sensitive,castellano2017relating,laurence2019spectral,gleeson2016effects,lago2000fast} to infer the network links from observations of nodal dynamics. We anticipate that our formalism will be useful in many different scenarios of technological and clinical significance involving time series data, including inference of neural, biochemical, genetic, communication, or social interaction networks and understanding their resilience to perturbations \cite{runge2015identifying} and predicting their critical behavior \cite{eroglu2020revealing}. Moreover, from a machine learning perspective, similar ideas could be important in constructing latent models of unknown environments solely from observations, and reconstructing unknown external inputs to networks \cite{kahl2019structural}. This may prove useful in optimal control theory or reinforcement learning of networks. 

We will test our network inference method on both simulated and experimental time series data from delay-coupled opto-electronic oscillator networks. An opto-electronic oscillator with time-delayed feedback is a dynamical system that can display a wide variety of complex behaviors, including periodic dynamics \cite{yao1996optoelectronic}, breathers \cite{kouomou2005chaotic}, and broadband chaos \cite{callan2010broadband}. Opto-electronic oscillators have found applications in highly stable microwave generation for frequency references \cite{yao1996optoelectronic}, neuromorphic computing \cite{larger2012photonic,paquot2012optoelectronic}, chaotic communications \cite{argyris2005chaos}, and sensing \cite{yao2017optoelectronic}. The nonlinear dynamics of individual \cite{larger2010optoelectronic,larger2013complexity,chembo2019optoelectronic} and coupled opto-electronic oscillators \cite{illing2011isochronal,ravoori2011robustness,williams2013experimental,hart2016experimental} are well-understood, making networks of opto-electronic oscillators an excellent test bed for network inference techniques. 

We find that our method accurately reconstructs the network from experimentally measured time series data, as long as the coupling is sufficiently strong and the network does not display strong global synchronization. We also find that the presence of dynamical noise, {\color{black} and heterogeneity of delays along network links,} may have a significant positive effect on the ability to infer links. Our results provide a clear demonstration that reservoir computing, and possibly other related machine learning methods, can provide accurate network inference for real networks, including situations where complications like noise and time delays in the coupling are present.

This paper is organized as follows. In Sec. 2, we introduce our network inference method for a general delay-coupled network dynamical system. In Sec. 3, we present the opto-electronic oscillator networks that we use for testing our method, along with a brief description of the collective dynamics of these networks in different parameter regimes. Section 4 presents results of our tests of the effectiveness of our method for both simulated and experimental time series data. Finally, we conclude in Sec. 5 with further discussion, suggested future directions, and possible generalizations of our method.

\section{Reservoir Computing Methodology for Network Inference}
\subsection{The General Delay-coupled Network}
In this section, we present the principles of our RC-based network link inference method. We consider a system of $D_n$ nodes, with the interactions among them mediated by a network of time-delayed links. {\color{black} For simplicity of presentation, in this section, we restrict ourselves to networks with identical delays along different links. Later in this paper (Sec. 4.3), we consider application of our framework to networks with heterogeneous link delays. For the present purpose, } we assume that the state of the $i$-th node in the network is given by a time dependent vector ${\mathbf {{X}}_{i}}[t]$ of dimension $D_{s}$, with $i=1,2,3,...,D_n$. We denote the components of this vector by $X_{i}^{\mu}$ with $\mu=1,2,3,...,D_s$. The coupled dynamics of the full system is governed by a general delay differential equation of the form
\begin{equation}\label{delay_genereic}
   \frac{dX_i^\mu(t)}{dt}=F_i^\mu\left[\mathbf{X}_i(t);\mathbf{X}_1(t-\tau),\mathbf{X}_2(t-\tau),...,\mathbf{X}_{D_n}(t-\tau);t\right].
\end{equation}
 Here ${{F_{i}^{\mu}}}$ is the function governing the dynamics of the $\mu$-th component of the state vector of the $i$-th node. ${{F_{i}^{\mu}}}$ is a function of $X_{j}^{\nu}$ if and only if there is a network link (connection) from the $\nu$-th component of the state vector of the $j$-th node to the $\mu$-th component of the state vector of the $i$-th node. Note that, as previously noted, for simplicity, in the above equation, we assume that the couplings have only a single time delay $\tau$, which is the time it takes a signal to propagate from one component of the system to another. In the experiments we shall consider here, the time series data from the above dynamics is sampled at a time interval $\Delta t={\tau}/{k}$ (with $k$ being an integer) and are denoted by $\left\{ {\bf {X}}_i\left[\Delta t\right]\right \},\left\{ {\bf {X}}_i\left[2\Delta t\right]\right \}, \left\{{\bf {X}}_i\left[3\Delta t\right]\right \}$, ... and so on. 

The problem we wish to address can be formalized as follows: \textit{If the observed time series data $\left \{{\bf {X}}_i[t]\right \}$ is the only information from the system we have, can we infer the connections of the network assuming that the underlying dynamical equations are of the general form as in Eq. \eqref{delay_genereic}?} We note that we lack any explicit knowledge of the functions $\left \{ {{F_{i}^{\mu}}} \right \}$. However, we shall henceforth assume that we know the delay $\tau=k\Delta t$, which, as shown in Appendix A, can, in the case of our opto-electronic test system, be inferred from the available time series data. Furthermore, we note that the performance of our method is not strongly dependent on the accuracy with which we infer the delay time $k$. For example, in our cases where we typically had $k=34$ (corresponding to a delay time $\tau=1.4$ms) in the simulations and experiments, setting the inferred value of $k$ anywhere between 34 and 37 (delay time of 1.4ms and 1.5ms, respectively) gave us essentially the same link inference results.  {\color{black}(We chose the delay time in order to work in a regime where the dynamics for our particular experimental test network is well-characterized  \cite{ravoori2011robustness,williams2013experimental,hart2016experimental,hart2015adding}). While the dynamics of the network depend on the delay time, we do not expect any change in the efficacy of our link inference technique for different delay times. In those cases, we only need to change the value of $k$ (as in Eqs. 4-9) in our inference procedures.} Finally, we note that in general situation, it might not be feasible to sample all the components of the state vectors. So, henceforth we will assume that we may sample only a subset of the components of each of the nodal state vectors $\left \{ {\bf {X}}_i[t] \right \}$, which, without loss of generality, we designate as the first $D_s'(\leq D_s)$ components. We now turn to a description of RC machine learning.
 
 \subsection{Time Series Prediction with a Reservoir Computer}
 
 \begin{figure}[h]\label{RC}
\includegraphics[scale=0.6,trim={1cm 0cm 0cm 0cm},clip]{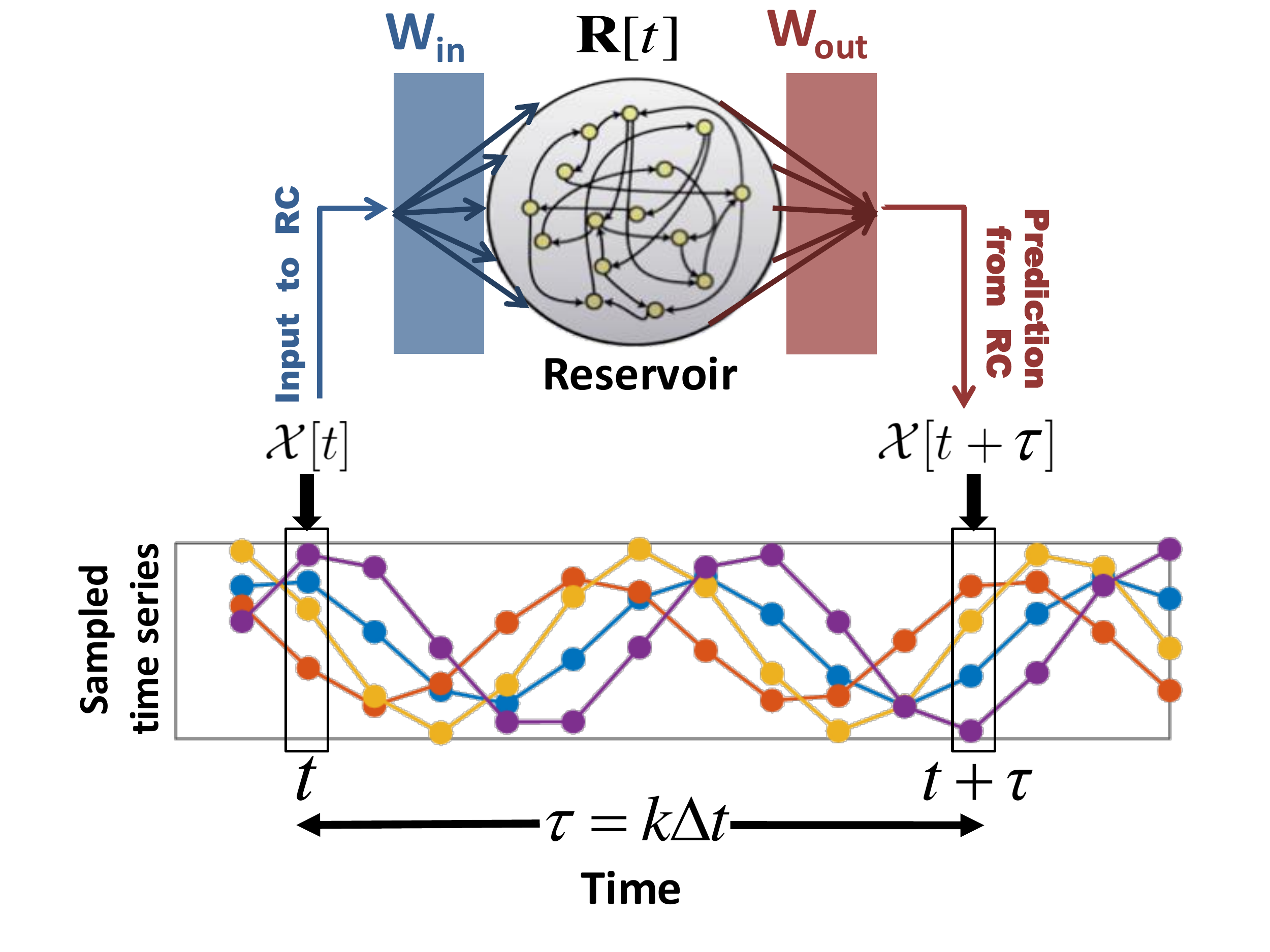}

\caption{Schematics of the reservoir computer (RC) trained for predicting the time series k time steps ahead. In the lower panel, the four time series represent scalar components of ${\bf{\mathcal{X}}}[t]$.}
\end{figure}

In this work, our first step is to train an RC to predict the time evolution of the node states one delay time $\tau=k\Delta t$ into the future, following which we will use that training information to extract the network structure of the system. A schematic of our RC implementation (\cite{banerjee2019using,pathak2018model,jaeger2004harnessing}) is shown in Fig. 1. We consider an RC network consisting of a large number of nodes $D_r$ (such that $D_r >> D_n \times D_s' \equiv D_X$). Each of the nodes has a time dependent scalar state, all of which are collected in a column vector ${\bf {R}}$ of length $D_r$. These reservoir nodes receive measured inputs of the unknown network system states $\left \{ {\bf {X}}_i[t] \right \}$. We concatenate the sampled input measurements of the time-dependent node state vectors $\left \{ {\bf {X}}_i[t] \right \}$ and place them in a single time-dependent column vector ${\bf{\mathcal{X}}}[t]$ of length $D_X$, such that the components of ${\bf{\mathcal{X}}}[t]$ are arranged as follows; 
\begin{equation}\label{x_states}
  {\bf{\mathcal{X}}}[t]=\left(X_1^1[t], X_1^2[t], ..., X_1^{D_s'}[t], X_2^1[t], X_2^2[t],..., X_2^{D_s'}[t],..., X_{D_n}^{D_s'}[t]\right)^T.  
\end{equation}
This vector is fed into the reservoir via a $D_r \times D_X$ input-to-reservoir coupling matrix ${\bf {W_{\text{in}}}}$ (Fig. 1). Furthermore, the reservoir nodes are coupled among themselves with a $D_r \times D_r$ adjacency matrix ${\bf{H}}$. The time evolution of the reservoir node states ${\bf {R}}$ are given by the equation,
\begin{equation}\label{RC_dynamics}
 {\bf{R}}\left[n\Delta t\right]=\sigma\left({\bf{HR}}\left[\left(n-1\right)\Delta t\right]+{\bf {W_{\text{in}}}{\mathcal{X}}}\left[n\Delta t\right]\right),  
\end{equation}
which maps the reservoir state at time $(n-1)\Delta t$ to the reservoir state at the next time step, $n\Delta t$, where $n$ is a positive integer, and $\sigma$ is a sigmoidal activation function acting componentwise on its vector argument (which has the same dimension, $D_r$, as ${\bf{R}}$). 

Keeping in mind the form of Eq. \eqref{delay_genereic}, our first step is to predict the future values of the sampled components of $ \left \{ {\bf{X}}_i\left[\left(n+k\right)\Delta t\right] \right \}$ in the concatenated form ${\bf{\mathcal{X}}}\left[\left(n+k\right)\Delta t\right]$ (Eq. \eqref{x_states}) from their current observed values ${\bf{\mathcal{X}}}\left[n\Delta t\right]$ using the reservoir state vector ${\bf{R}}\left[n\Delta t\right]$. In our case, this is done by using a suitable linear combination of the reservoir node states with a $D_X \times D_r$ reservoir-to-output coupling matrix ${\bf {W_{\text{out}}}}$  (Fig. 1) according to the equation,
\begin{equation}\label{w_out}
 {\bf{\mathcal{X}}}^P\left[\left(n+k\right)\Delta t\right]= {\bf {W_{\text{out}}}}  {\bf{R}}\left[n\Delta t\right],
\end{equation}
where the superscript $P$ indicates that the vector is a prediction from the RC, as opposed to being sampled from the actual system. During the training time, we measure the system training time series data $\left \{ {\bf{\mathcal{X}}}\left[t\right] \right \}$ from the unknown system of interest for a large number of time steps. We use this data along with Eq. \eqref{RC_dynamics} to generate the time series data for the RC nodal states $\left \{{\bf{R}}\left[n\Delta t\right]\right \}$, which we store. For the training of the RC, we then find the elements of the matrix ${\bf {W_{\text{out}}}}$ by doing a linear regression from these stored reservoir states ${\bf{R}}\left[n\Delta t\right]$ to the measured time-advanced system states $\left \{ {\bf{\mathcal{X}}}\left[\left(n+k\right)\Delta t\right] \right \}$, such that Eq. \eqref{w_out} provides a best mean squared fit of the prediction ${\bf{\mathcal{X}}}^P\left[\left(n+k\right)\Delta t\right]$ to the measured state ${\bf{\mathcal{X}}}\left[\left(n+k\right)\Delta t\right]$. This amounts to the minimization of a cost function ${\mathcal {C}}$ given by
\begin{equation}\label{cost}
 {\mathcal {C}}= \sum_{{\text{Training Steps n}}} \left|\left |{{\bf{\mathcal{X}}}}\left[\left(n+k\right)\Delta t\right]- {\bf {W_{\text{out}}}}  {\bf{R}}\left[n\Delta t\right]\right|\right |^{2}+\lambda\left|\left|{\bf {W_{\text{out}}}}\right|\right|^{2}
\end{equation}
where the last term ($\lambda\left|\left|{\bf {W_{\text{out}}}}\right|\right|^{2}$) is a ``ridge" regularization term \cite{hoerl1970ridge} used to prevent overfitting to the training data and $\lambda$ is typically a small number. 

To completely specify the training procedure that we use, we now specify the structures of the different associated matrices. The elements of the input matrix ${\bf {W_{\text{in}}}}$ are chosen randomly from a uniform distribution in the interval $\left[-w,w\right]$. The reservoir connectivity matrix ${\bf{H}}$ is a sparse random matrix, corresponding to an average in-degree $d_{\text{av}}$ of the reservoir nodes. The non-zero elements of ${\bf{H}}$ are chosen randomly from an uniform distribution $\left[-h,h\right]$ and $h$ is chosen such that the spectral radius of ${\bf{H}}$ (i.e., the maximum magnitude of the eigenvalues of ${\bf{H}}$) is equal to some predefined value $\rho$. The hyperparameters $w$ and $d_{\text{av}}$ are chosen using a Nelder-Mead optimization procedure where we minimize ${\mathcal {C}}$ for a representative training data and the corresponding output matrix ${\bf {W_{\text{out}}}}$ found from the training data. [We use $w=1.17$ and $d_{\text{av}}=2.38$ for tests on simulated data and $w=1.19$ and $d_{\text{av}}=2.56$ for tests on experimental data. We typically use values $\lambda=10^{-4}$ and $\rho=0.9$ for the other two hyperparameters, $3\times10^4$ steps (about 880 delay times) for training, and the sigmoidal activation function $\sigma$ is taken to be the hyperbolic tangent function. The reservoirs we used typically had $3000$ nodes]. After a successful training of the RC with these specifications, Eq. \eqref{w_out} can be seen as an \textit{in silico} model for the dynamics of the actual system. Explanations for the special properties of trained RCs, which allow us to use RCs for our purpose, can be found in Refs. \cite{pathak2017using,lymburn2019reservoir,lu2018attractor,bollt2021explaining}. 

%Finally, we mention that we expect our method to work irrespective of the microscopic details of the dynamics or the physical implementation of the reservoir computer\cite{tanaka2019recent,chembo2020machine,markovic2020physics}, e.g., with photonic \cite{larger2017high,vinckier2015high,duport2016fully}, electronic and opto-electronic \cite{schrauwen2008compact,appeltant2011information}, neuronal \cite{enel2016reservoir}, molecular \cite{goudarzi2013dna}, or chemical \cite{nguyen2020reservoir} systems.

\subsection{Our Network Inference Procedure}
We now describe how we use the training results of the previous subsection to obtain the network structure of the unknown system. We first briefly discuss how the form of Eq. \eqref{delay_genereic} allows us to relate the network structure to the spread of the effect of small perturbations to the system. Suppose that, at a time point $t=n\Delta t$, we perturb the $\nu$-th component of the state of node $j$ by an infinitesimal amount $\delta X_j^{\nu}\left[n\Delta t\right]$. Differentiating both sides of Eq. \eqref{delay_genereic}, we see that this perturbation changes the $\mu$-th component of the state of node $i$($\neq j$) at a later time $(n+k)\Delta t=t+\tau$ via the corresponding change in the time derivative, 
\begin{equation}\label{delay_perturbed}
   \delta \left( \frac{dX_i^{\mu}}{dt}\Bigg|_{t+\tau}\right)= \frac{\partial F_i^{\mu}}{\partial X_j^{\nu}}\delta X_j^{\nu}\left[t\right]+\mathcal{O}\left(\left(\delta X_j^{\nu}\left[t\right]\right)^{2}\right).
\end{equation}
 This equation shows that, to lowest order, the effect of the small perturbation on component $\nu$ of the state of node $j$ is propagated to the component $\mu$ of the state of node $i$ with a delay of $\tau=k\Delta t$ provided that there is a corresponding network link between them, i.e., if ${\partial F_i^{\mu}}/{\partial X_j^{\nu}} \neq 0$. In particular, propagation of a perturbation from component $\nu$ of the state of node $j$ to component $\mu$ of the state of node $i$ with a delay of $k$ time steps implies that a directional network link, $(j,\nu)\rightarrow(i,\mu)$, exists between them.

While the above discussion is predicated on application of an active perturbation, we see that the result is essentially determined by the partial derivative ${\partial F_i^{\mu}}/{\partial X_j^{\nu}}$. Thus we wish to determine whether this partial derivative is zero (corresponding to the absence of a link) or not (corresponding to the presence of a link). We attempt to do this by use of the trained RC (which, as we emphasize, was obtained solely from observations, i.e., non-invasively). Indeed, when Eq. \eqref{w_out} is approximately true for a well-trained RC with ${\bf{\mathcal{X}}}^P\left[\left(n+k\right)\Delta t\right] \approx {\bf{\mathcal{X}}}\left[\left(n+k\right)\Delta t\right]$, we can use that equation to consider the RC-predicted dynamics as a proxy for the dynamics of the actual system. In that case, within this assumed RC proxy model, we can analytically assess the effects of small perturbations, and compare them to Eq. \eqref{delay_perturbed}. To do so, we first combine Eqs. \eqref{RC_dynamics} and \eqref{w_out} for the RC, and assume the relation ${\bf{\mathcal{X}}}^P\left[\left(n+k\right)\Delta t\right] = {\bf{\mathcal{X}}}\left[\left(n+k\right)\Delta t\right]$ for the training data to obtain the equation,
\begin{equation}\label{RC_combined}
    {\bf{\mathcal{X}}}\left[\left(n+k\right)\Delta t\right]= {\bf {W_{\text{out}}}}  \sigma\left({\bf{HR}}\left[\left(n-1\right)\Delta t\right]+{\bf {W_{\text{in}}}{\mathcal {X}}}\left[n\Delta t\right]\right),
\end{equation}
where the time points belong to the training time series. In order to evaluate the effect of a perturbation to one node on another, we desire to eliminate reservoir variables ${\bf{R}}$ from this equation. Naively, this could be done by solving Eq. \eqref{w_out} for ${\bf{R}}\left[\left(n-1\right)\Delta t\right]$ in terms of  ${\bf{{\mathcal{X}}}}\left[\left(n+k-1\right)\Delta t\right]$. However, the number of components of ${\bf{R}}$ is large compared to the number of components of ${\bf{{\mathcal{X}}}}$, and so there are many solutions of Eq. \eqref{w_out} for ${\bf{R}}$. As in our previous work \cite{banerjee2019using}, we hypothesize (and subsequently test) that, for our purpose, the Moore-Penrose pseudoinverse \cite{penrose1955generalized} (symbolically denoted by ${\bf \hat{W}_{\text{out}}}^{-1}$) provides a useful solution of the equation ${\bf{\mathcal{X}}}\left[\left(n+k\right)\Delta t\right]={\bf {W_{\text{out}}}}  {\bf{R}}\left[\left(n-1\right)\Delta t\right]$ for ${\bf{R}}\left[\left(n-1\right)\Delta t\right]$. With this, Eq. \eqref{RC_combined} becomes
\begin{equation}\label{RC_combined_2}
 {\bf{{\mathcal{X}}}}\left[\left(n+k\right)\Delta t\right]= {\bf {W_{\text{out}}}}  \sigma\left({\bf {H{\hat W}_{\text{out}}}}^{-1}{\bf{{\mathcal{X}}}}\left[\left(n+k-1\right)\Delta t\right]+{\bf {W_{\text{in}}}{\mathcal{X}}}\left[n\Delta t\right]\right)   
\end{equation}
yielding a putative dynamical model for the system from which we will now study the effect of smal perturbations. Thus, an infinitesimal amount of change in the network node states at time step $n\Delta t$, written as $\delta {\bf {{\mathcal{X}}}}\left[n\Delta t\right]$,  propagates to a change at time $(n+k)\Delta t$ as described by differentiating Eq. \eqref{RC_combined_2},
\begin{equation}\label{perturbed_RC1}
\delta{\bf{{\mathcal{X}}}}\left[\left(n+k\right)\Delta t\right]=\left(\mathbb{1} -  {\bf {W_{\text{out}}}}{\bf {\hat{H}}}\left[n\Delta t\right]{\bf{\hat W}_{\text{out}}}^{-1}\right)^{-1}{\bf {W_{\text{out}}}}{\bf {{\hat {W}}_{\text{in}}}}\left[n\Delta t\right]\delta{\bf {{\mathcal{X}}}}\left[n\Delta t\right]\\
\equiv {\bf {M}}\left[n\Delta t\right]\delta{\bf {{\mathcal{X}}}}\left[n\Delta t\right],  
\end{equation}
where we have used Eq. \eqref{RC_dynamics}, assumed that $\Delta t$ is sufficiently small that $\delta {\bf {{\mathcal{X}}}}\left[(n+k)\Delta t\right] \approx \delta {\bf {{\mathcal{X}}}}\left[(n+k-1)\Delta t\right]$, and defined the new matrix elements ${{\hat{H}}}_{ij}\left[n\Delta t\right]=H_{ij}\sigma'\left(R_i[n\Delta t]\right)$ and $\left({{\hat{W}}}_{\text{in}}\right)_{ij}\left[n\Delta t\right]=\left(W_{\text{in}}\right)_{ij}\sigma'\left(R_i[n\Delta t]\right)$ where $\sigma'(u)=d\sigma(u)/du$. We now employ this equation in component form in the place of Eq. \eqref{delay_perturbed} as a proxy approximating how small perturbations spread across the network. In particular, just like the partial derivative ${\partial F_i^{\mu}}/{\partial X_j^{\nu}}$ determines whether a change at the $\nu$-th component of state of node $j$ results in a change of the $\mu$-th component of the state of node $i$ after $k$ time steps in Eq. \eqref{delay_perturbed}, $M_{i,j}[n\Delta t]$ determines the same in Eq. \eqref{perturbed_RC1}, when used in conjunction with our definition Eq. \eqref{x_states} for $ {\bf {{\mathcal{X}}}}\left[ t\right]$. 

We now describe how we use our determination of ${\bf {M}}$ in Eq. \eqref{perturbed_RC1} to recover the network structure. For this purpose, {\color{black} based on our knowledge of $M_{i,j}[n\Delta t]$,} we are interested only in determining whether ${\partial F_i^{\mu}}/{\partial X_j^{\nu}}$ is zero (no link $(j,\nu)\rightarrow(i,\mu)$) or not. {\color{black}In the true} Jacobian, the absence of link would imply ${\partial F_i^{\mu}}/{\partial X_j^{\nu}}=0$ exactly. However, in our procedure, there are errors and thus, {\color{black} the elements of ${\bf {M}}\left[n\Delta t\right]$} are never zero. These errors are due to finite reservoir size, finite training data length, noise in the training data, and the Moore-Penrose inversion which, as we hypothesized, is useful but not exact.

{\color{black} Generally speaking, in past link inference methods, the common approach is to somehow obtain a {\color{black}time-independent,} continuous-valued score $S_{ij}$ hopefully accurately  reflecting the likelihood that node $i$ is linked to $j$. Once the score is found, as is explained below, one can then choose an appropriate statistical technique for translating the score into a good binary choice of whether or not $i\rightarrow j$ corresponds to an actual link. In essence, the key goal of our paper is to use machine learning to obtain good scores, and for that purpose we will use our machine learning determination of ${\bf {M}}$.}

 To form an appropriate score {\color{black} corresponding to} each of the {\color{black}time-dependent} elements $M_{ij}$, we use $S_{ij}\equiv \left\langle\left|M_{ij}\right|\right\rangle_t$ where $\langle\rangle_t$ denotes time-averaging over a  time {\color{black} sufficiently long that the scores $S_{ij}$ do not change appreciably upon, e.g., doubling the averaging time}. In our tests (Sec. 4), the averaging time is 1000$\Delta t$. Here the absolute value of $M_{ij}$ is to be taken so that the positive and negative values do not cancel each other while doing the time averaging. {\color{black} If this assigned score $S_{ij}$ is above a threshold for a particular element $M_{ij}$, we assume that there is a network link corresponding to that element, and if the score is less than a threshold, we assume that the corresponding network link is absent. } 

With the scores $S_{ij}$ defined and calculated, choosing the threshold is a well-known problem of binary categorization of a collection of continuous numerical scores. {\color{black}Since obtaining a useful score for the network link inference is the goal of our machine-learning-based methodology, we regard a detailed discussion of thresholding or other follow-up statistical analysis of the obtained scores to be beyond the scope of this work. We comment only that once the scores $S_{ij}$ are found, one can then choose an appropriate statistical technique for translating the score into a good binary choice. This choice will depend on circumstances that are specific to the situation at hand (e.g., the cost of false positive link choice versus the cost of a false negative link choice).} This is a basic problem in statistics and addressed extensively in earlier works with methods such as Receiver Operating Characteristic curves \cite{fawcett2006introduction,yang2015evaluating}, Precision-Recall curves \cite{fawcett2006introduction,yang2015evaluating}, fitting to mixtures of statistical distributions \cite{fawcett2006introduction}, Bayesian techniques \cite{fawcett2006introduction}, etc. {\color{black} A recent paper \cite{cecchini2021impact} has proposed a binary classification technique specifically designed for network inference purposes.}

To avoid the details of the statistical methodologies, we adopt a procedure which is simple but sufficient for the purpose of evaluating the goodness of the scores of candidate links resulting from our link inference method. Thus we shall henceforth assume that we know the total number of links (denoted by $L$) in the unknown network and shall designate the largest $L$ scores $S_{ij}$ as corresponding to inferred network links, while those $M_{ij}$ with scores below the largest $L$ scores will be inferred to not correspond to network links. The performance of this link inference technique will be measured by the corresponding number of falsely
inferred links (“false positives”). Since we assume that we know the total number of links, the number of falsely inferred links is also equal to the number of missed links (“false negatives”). As we will show below, this method, applied to our machine-learning-determined scores, can produce excellent results in link inference tasks over a wide range of coupling strengths, network topologies, and noise levels. {\color{black} While in practice, $L$ may be unknown, and, depending on the situation, a user may wish to employ an appropriate statistical technique for making the above binary choice from our score, we claim that the results we obtained with $L$ assumed known indicate the value of our technique for score determination, without having to introduce and discuss more involved methods and their appropriateness in different situations (e.g., precision-recall is more appropriate for sparse networks than Receiver Operating Characteristic \cite{yang2015evaluating}). Later in this paper, at the end of Sec. 4.1, we shall present results that support this claim.}

\section{Opto-electronic Oscillator Networks}
\subsection{Description of the experiment}
In this section, we introduce our opto-electronic network used as a platform to test our network inference procedure.  An individual opto-electronic oscillator is a nonlinear, time-delayed feedback loop. Our network consists of four nominally identical opto-electronic oscillators with time-delayed optical coupling between them. The individual and coupled dynamics of opto-electronic oscillators have been studied extensively \cite{murphy2010complex,callan2010broadband,ravoori2011robustness,williams2013experimental,larger2013complexity,hart2016experimental,hart2019delayed,chembo2019optoelectronic}.

%For ease of implementation, we control $\tau^f$ using a delay buffer in our DSP board.}

{\color{black} An opto-electronic oscillator essentially consists of a nonlinearity whose output is fed back into its input with some feedback time delay $\tau^f$. This feedback delay $\tau^f$ is inherent to the oscillator, and without it, the system would have no dynamical behavior. A description of one of our opto-electronic oscillators follows:

A fiber-coupled continuous-wave laser emits light of constant intensity. The light passes through an intensity modulator, which serves as the nonlinearity in the feedback loop and can be modeled by $\cos^2(\pi v/2V_\pi)$ where $v$ is the voltage applied to the modulator. For our modulators $V_\pi=3.4$V. The feedback optical signal is converted to an electrical signal by a photodiode, which is then delayed and filtered by a digital signal processing (DSP) board (Texas Instruments TMS320C6713).  This signal is output by the DSP board, amplified, and fed back to drive the modulator. The normalized voltage $x(t) \equiv {\pi v(t)}/{2V_\pi}$ applied to the intensity modulator is measured and is our dynamical variable. If there were no DSP board, the delay would be controlled by the optical fiber length and the filtering would be done by the analog electronic components, such as the amplifier. The DSP board simply provides enhanced control over the delay and filter parameters.}

{\color{black}In general, when one couples two oscillators together, a coupling delay will be induced by the finite propagation time of the signal. That coupling delay becomes important when it is not too much shorter than the fastest system time scale. This is the case for our network of coupled opto-electronic oscillators. In general, for a network of oscillators with $L$ links, there will be $L$ different coupling delays. We use the notation $\tau^c_{ij}$ to refer to the coupling delay in the link from node $j$ to node $i$. In our experiments, we choose all the coupling delays to be identical, such that $\tau^c_{ij}=\tau^c$ (heterogeneous delays are considered in Sec. 4.3).

\begin{figure}
    \centering
    \includegraphics[width=0.9\textwidth]{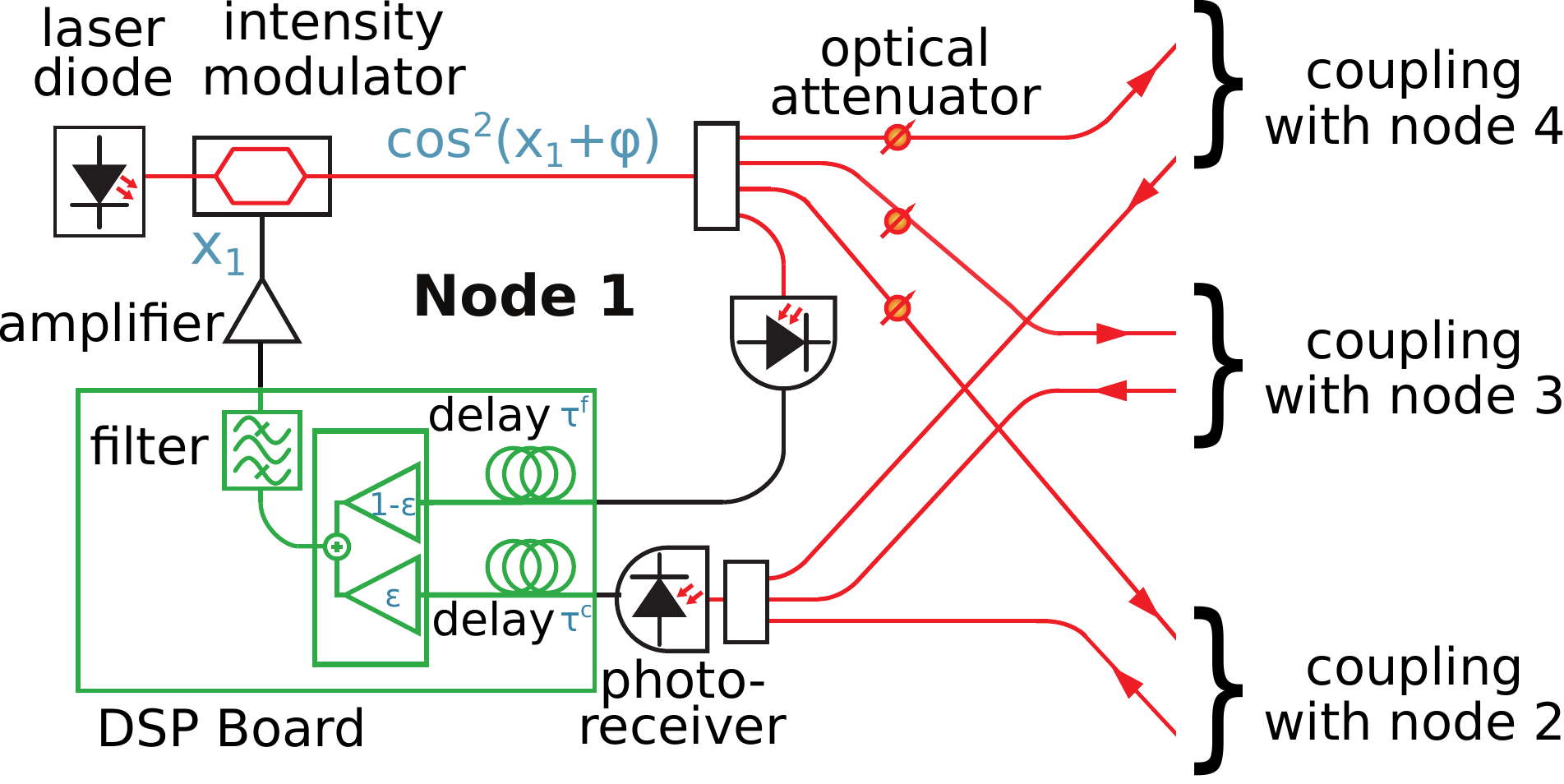}
    \caption{Illustration of opto-electronic oscillator and coupling scheme. Red lines indicate signal paths in optical fibers. Black lines are used to indicate electronic signals, and green indicates the digital signal processing (DSP) board.}
    \label{fig:OEO}
\end{figure}

An illustration of a single networked opto-electronic oscillator is shown in Fig. 2. In order to couple the opto-electronic oscillator to other nodes in the network, a $1\times 4$ fiber optic splitter is introduced after the intensity modulator. One of the optical outputs of the splitter is fed back into that node's DSP board, creating a self-feedback electrical signal. The other three splitter outputs are sent to the other three nodes in the network. Incoming optical signals from the other nodes pass through variable optical attenuators, which control the link strengths, then are combined by a 3x1 fiber optic combiner. This combined optical signal is converted into a coupling electrical signal by a second photodiode. 

The feedback and coupling signals are delayed independently in the DSP board such that $\tau^f$ and $\tau^c$ represent the feedback and coupling delay times, respectively. The outputs of the two delay lines are weighted and combined in the DSP board. This combined signal is output by the DSP, amplified, and used to drive the modulator. The ratio of amplification factors of the coupling signal and feedback signal is given by the coupling strength $\epsilon$.}

The amplifier gain is set such that each feedback loop has identical round-trip gain $\beta=3.8$, phase bias $\phi_0=\pi/4$, and feedback time delay $\tau^f$ = 1.4 ms such that a single uncoupled node behaves chaotically. The digital filter implemented by the DSP board is a two-pole Butterworth bandpass filter with cutoff frequencies $\omega_H/2\pi$ = 100 Hz and $\omega_L/2\pi$ = 2.5 kHz and a sampling rate of 24 kSamples/s. {\color{black} These parameters were chosen because the experimental system with these parameters has been well-characterized \cite{ravoori2011robustness,williams2013experimental,hart2015adding,hart2016experimental}.}

For each set of measurements, the nodes are initialized from noise from the electrical signal into the digital signal processing (DSP) board.
Then feedback is turned on without coupling, and the opto-electronic oscillators are allowed to oscillate independently until transients die out. At the end of this period, the coupling to the other nodes is turned on and the voltage reading $x(t)$ of each opto-electronic oscillator is recorded on an oscilloscope.

%Figure 3 shows some examples of the dynamics displayed by our network of opto-electronic oscillators. The upper left time series is a measurement of the four opto-electronic oscillators arranged in the six-link network shown. These dynamics are synchronized and are also the dynamics of an individual uncoupled opto-electronic oscillator, since the effect of the Laplacian coupling vanishes for global synchronization. The lower left time series is a measurement of the four opto-electronic oscillators coupled in the nine-link network shown. In this case, the opto-electronic oscillators do not synchronize even though the coupling is strong ($\epsilon=0.6$).

\subsection{Mathematical Model and Numerical Simulations of the Opto-Electronic Network}
The equations governing the dynamics of our network of opto-electronic oscillators are derived in Ref. \cite{murphy2010complex} and are given by

%\begin{align}
%    \dot{x}_i(t) &= -(\omega_L+\omega_H)x_i(t) -\omega_Ly(t)- \omega_L\beta\cos^2\bigg(x_i(t-\Delta t)+\epsilon\sum_jA_{ij}\big(x_j(t-\Delta t)-x_i(t-\Delta t)\big)+\phi_0\bigg) \\
%    \dot{y}_i(t) &= \omega_Hx_i(t)
%\end{align}

%--OR--

%\begin{comment}

%\begin{equation}\label{eq_u}
%    \frac{d\mathbf{u}_i(t)}{dt}=\mathbf{Eu}_i(t)-\mathbf{F}\beta\cos^2\big(x_i(t)+\phi_0\big)
%\end{equation}

%\begin{equation}\label{eq_x}
%    \frac{dx_i(t)}{dt}=\mathbf{G}\left(\frac{d\mathbf{u}_i(t-\tau)}{dt}+\epsilon\sum_jA_{ij}\left(\frac{d\mathbf{u}_j(t-\tau)}{dt}-\frac{d\mathbf{u}_i(t-\tau)}{dt}\right)\right)+\xi_i(t)
%\end{equation}
%
%\end{comment}

{\color{black}

\begin{equation}\label{eq_u}
\begin{split}
    \frac{d\mathbf{X}_i(t)}{dt}=&\mathbf{EX}_i(t)-\beta\mathbf{G}\cos^2\left(X_i^{1}(t-\tau^f)+\phi_0\right)  \\  &-\epsilon\beta\mathbf{G}\sum_j A_{ij}[\cos^2\left(X_j^{1}(t-\tau^c_{ij})+\phi_0\right)-\cos^2\left(X_i^{1}(t-\tau^f)+\phi_0\right)]+{\mathbf{\xi}}_i(t)
    \end{split}
\end{equation}
}
where
\begin{equation}
\label{eq:u_mats}
    \mathbf{E}=\begin{bmatrix}-(\omega_L+\omega_H) & -\omega_L \\ \omega_L & 0\end{bmatrix}, \qquad
    \mathbf{G}=\begin{bmatrix}\omega_L\\0 \end{bmatrix},
\end{equation}

Here ${\mathbf{{X}}}_i=\left [X_i^1(t), X_i^2(t)\right ]^{T}$ (corresponding to $D_s=2$ in Eq. \eqref{x_states}) is the state of the digital filter of node $i$ (with $i,j\in\{1,2,3,4\}$, corresponding to $D_n=4$). By virtue of the second component of $\mathbf{G}$ being zero, coupling between nodes occurs only between $X^1_i$ and $X^1_j (i\neq j)$, where $X^1_i(t)$, the normalized voltage of the electrical input to the intensity modulator and is also the only observed variable (i.e., $X^1_i(t)=x(t)$, corresponding to $D_s'=1$). {\color{black}The nodes are coupled via the adjacency matrix $A_{ij}$, such that $A_{ij}=1$ if there is a link to the first component of the state vector of node $i$ from the first component of the state vector of node $j$, and $A_{ij}=0$ otherwise.} Since the coupling is between only the first components of the vectors ${\mathbf{X}}_i$, we have dropped the component indices in the adjacency matrix in Eq. \eqref{eq_u}. The coupling strength is given by $\epsilon$, and $\mathbf{E}$ and $\mathbf{G}$ are matrices that describe the filter. Finally, ${\mathbf{\xi}}_i(t)=\left [\xi_i^1(t), \xi_i^2(t)\right ]^{T}$ is a vector corresponding to white noise acting independently at each oscillator, and its components have the property that $\left\langle \xi_i^{\mu}[s] \xi_j^{\nu}[t]\right\rangle=2\zeta\delta(s-t)\delta_{ij}\delta_{\mu \nu}$ with $\zeta$ denoting the strength of the noise.

{\color{black}
In our experiments, we choose all the feedback delays and coupling delays to be nominally equal (i.e., $\tau^f=\tau^c_{ij}=\tau$). In this case, Eq. \ref{eq_u} describes a network with Laplacian coupling:

\begin{equation}\label{eq_u_Laplacian}
    \frac{d\mathbf{X}_i(t)}{dt}=\mathbf{EX}_i(t)-\beta\mathbf{G}\cos^2\left(X_i^{1}(t-\tau)+\phi_0\right)-\epsilon\beta\mathbf{G}\sum_j L_{ij}\cos^2\left(X_j^{1}(t-\tau)+\phi_0\right)+{\mathbf{\xi}}_i(t)
\end{equation}
}
In this case the nodes are coupled via the Laplacian connectivity matrix $L_{ij}$, defined so that $L_{ij}=1$ if there is a link to the first component of the state vector of node $i$ from the first component of the state vector of node $j$, $L_{ij}=0$ if there is no such link, and $L_{ii}=-\sum_{j\neq i}L_{ij}$. {\color{black}Laplacian coupling tends to lead to global synchronization, which is a particularly challenging case for link inference, as we show in the following sections}.

Since the coupling is between only the first components of the vectors ${\mathbf{X}}_i$, we have dropped the component indices in the Laplacian adjacency matrix in Eq. \eqref{eq_u}. Comparison with Eq. \ref{delay_genereic} shows that in our example,  $\boldsymbol{F}_i=\mathbf{EX}_i(t)-\beta\mathbf{G}\cos^2\left(X_i^{1}(t-\tau)+\phi_0\right)-\epsilon\beta\mathbf{G}\sum_j L_{ij}\cos^2\left(X_j^{1}(t-\tau)+\phi_0\right)+{\mathbf{\xi}}_i(t)$, where we have dropped the component superscripts. The oscillators are identical, so these functions are independent of $i$, except for the noise term. The relevant partial derivative controlling the propagation of perturbation is ${\partial F_i^{\mu}}/{\partial X_j^{\nu}} \propto L_{ij}\delta_{\mu 1}\delta_{\nu 1}$, for $i\neq j$.

While Eq. \eqref{eq_u} accurately describes the behavior of our network of opto-electronic oscillators, numerical simulations are inherently discrete in time. Instead of discretizing Eq. \eqref{eq_u} directly, our simulations use a discrete-time model based on the discrete-time filter equations implemented on the DSP board, which can be found in Ref. \cite{hart2018experiments} and is explained in Appendix B. In particular, for this case, we characterize the noise strength by the variable $\kappa$, so that $\kappa=\zeta \Delta t$. 
%In the numerical simulations, we initialize the network states ($X_i^1, i=1,2,3,4$) over the duration of delay time with white noise having mean $0.0256$ and standard deviation $0.0394$, uncorrelated over time and nodes. Then we set the coupling strength $\epsilon=0$ and simulate the dynamics of the uncoupled nodes for $0.384$s. After that we simulate the dynamics of the full coupled system. 
For the discrete equation that we simulate, the time step is $0.04$ms, which corresponds to the $2.4 \times 10^4$ samples/second  sampling rate used by the digital filter in our experiment.

\begin{figure}[h]\label{sync_ex}
\includegraphics[scale=0.33,trim={1cm 0cm 0cm 0cm},clip]{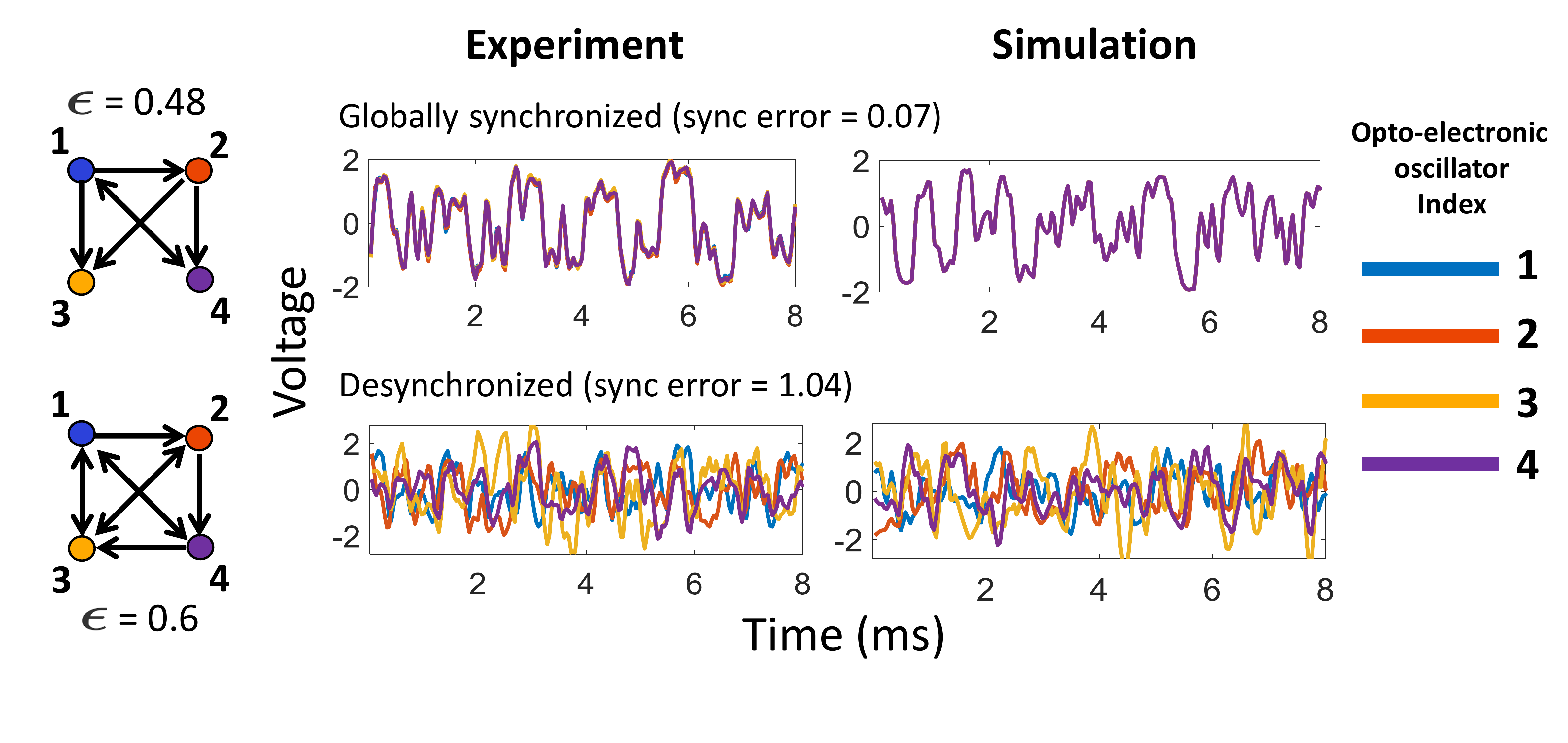}
\caption{Examples of experimental and simulated time series from two opto-electronic oscillator networks showing globally synchronous (upper panels) and completely desynchronized (lower panels) behavior respectively. For the simulations, we have used the noise strength $\kappa=10^{-6}$. In the plots, purple overlays orange, which overlays red, which overlays blue.}
\end{figure}

Our model is verified by comparison with the experiment, as shown in Fig. 3 for two sets of examples. The upper panel shows experimentally measured (left) and simulated (right) time series of the four opto-electronic oscillators arranged in the six-link network shown. The dynamics are synchronized and are also the dynamics of an individual uncoupled opto-electronic oscillator, since the effect of the Laplacian coupling vanishes for global synchronization. The lower panel shows a measurement (left) and simulation (right) of the four opto-electronic oscillators coupled in the nine-link network shown. In this case, the opto-electronic oscillators do not synchronize even though the coupling is strong ($\epsilon=0.6$). In both cases, the simulations are in good agreement with the experiment

As we shall see, the degree of synchronization of the oscillators in the network is an important factor in the success of our method to infer the network topology. In order to quantify the degree of global synchrony, we define synchronization error as 

\begin{equation}
    \label{eq:syncerror}
    \textrm{Synchronization Error} = \frac{1}{D_n(D_n-1)}\left\langle\sum_{i,j}|x_i(t)-x_j(t)|\right\rangle_t
\end{equation}

 where $\langle\rangle_{t}$ means time average over a sufficiently long time. This non-negative measure decreases with the amount of synchronization in the system and is zero for perfect global synchrony. For example, in Fig. 3, the synchronized examples (upper panel) have synchronization error $\approx 0.07$, whereas the desynchronized examples (lower panel) have synchronization error $\approx 1.04$.

\begin{figure}[h]\label{sync_err}
\includegraphics[scale=0.6,trim={0cm 6cm 0cm 0cm},clip]{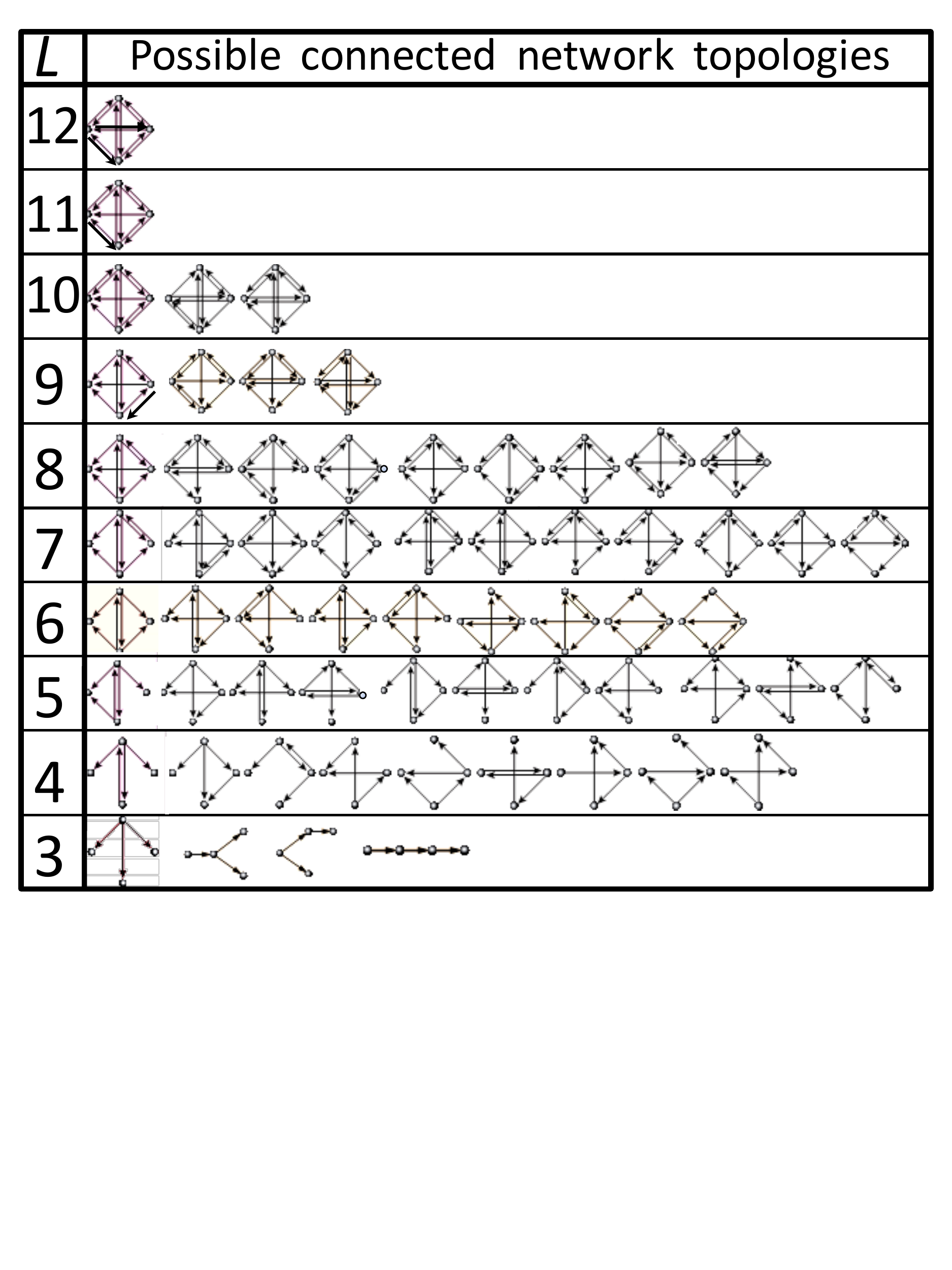}\\
\caption{ List of possible connected directed 4-node networks \cite{ravoori2011robustness} with different numbers of links ($L$).}
\end{figure}

Using computer simulations, we have studied the dependence of the synchronization error on the network coupling strength $\epsilon$ and the number $L$ of network links for all possible directed and connected networks with 4 opto-electronic oscillator nodes. The list of the $62$ possible networks is shown in Fig. 4, and is adapted from Ref. \cite{ravoori2011robustness}. Figure 5 shows the synchronization behavior of these networks as a function of the coupling strength $\epsilon$ for fixed $\beta=3.8$ and $\phi_0=\pi/4$. In Fig. 5, the color coded synchronization error for each of the $62$ networks in Fig. 4 is shown as one of the 62 horizontal bars for each value of the coupling strength. Here, the convention we follow is that, for fixed number of links ($L$), moving upwards, the horizontal bars correspond to the networks listed in Fig. 4 left to right. The same convention is followed in Figs. 6, 7 and 10. The results in Fig. 5 were obtained from numerical simulations without noise ($\kappa=0$). We see that for intermediate coupling strengths, the networks synchronize, but for small and large coupling strengths, the networks do not synchronize. The seemingly counterintuitive behavior that large coupling strengths lead to desynchronization has been studied for our network of opto-electronic oscillators \cite{hart2015adding} and is characteristic of delay coupled systems in general \cite{flunkert2010synchronizing}. Furthermore, for coupling strengths in the range $\epsilon >0.5$, for a fixed value of coupling strength, sparser networks are seen to synchronize more readily than densely connected networks. This behavior is also studied and explained in earlier works \cite{hart2015adding, townsend2020dense}.

\begin{figure}[h]\label{sync_err}
\includegraphics[scale=0.5,trim={0cm 0cm 0cm 0cm},clip]{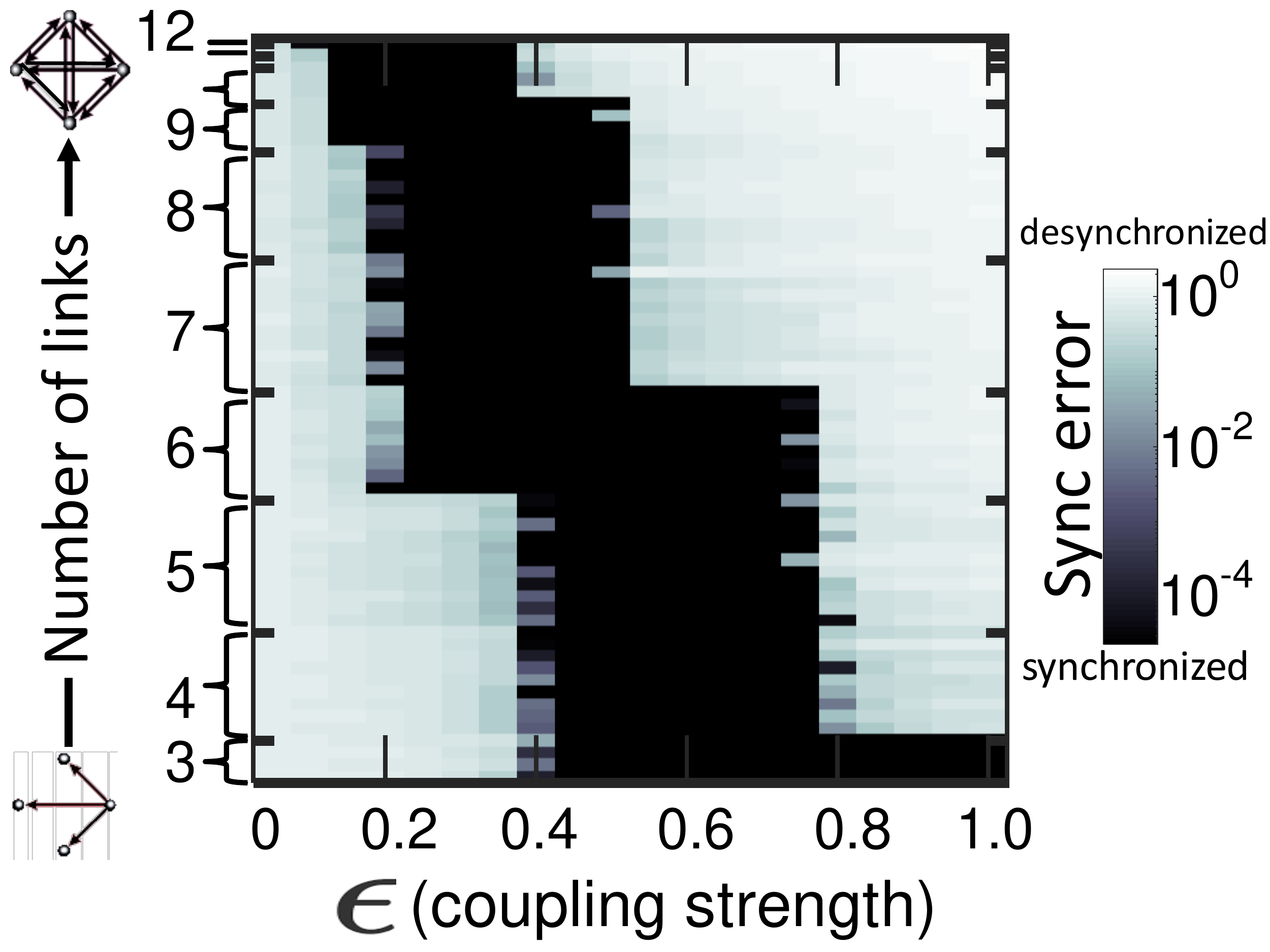}
\caption{ Synchronization error for simulated time series of the networks in Fig. 4 realized with opto-electronic oscillator nodes with different coupling strengths for random initial conditions.The color coded synchronization error for each of the $62$ networks in Fig. 4 is shown as one of the $62$ horizontal bars for each value of the coupling strength. Here, the convention we follow is that, for fixed number of links ($L$), moving upwards, the horizontal bars correspond to the networks listed in Fig. 4 left to right. The same convention is followed in Figs. 6, 7 and 10.}
\end{figure}

\section{Results of Link Inference Tests}
In this section, we present tests of the efficacy of our machine learning technique. These include numerical simulation tests for simulations with homogeneous link delays (Sec. 4.1), opto-electronic experimental tests with nominally homogeneous link delays (Sec. 4.2), and numerical simulation tests with inhomogeneous link delays (Sec. 4.3).

\begin{figure}[!ht]\label{increasing_noise}
{\bf(a)}\\
\includegraphics[scale=0.4,trim={0cm 0cm 0cm 0cm},clip]{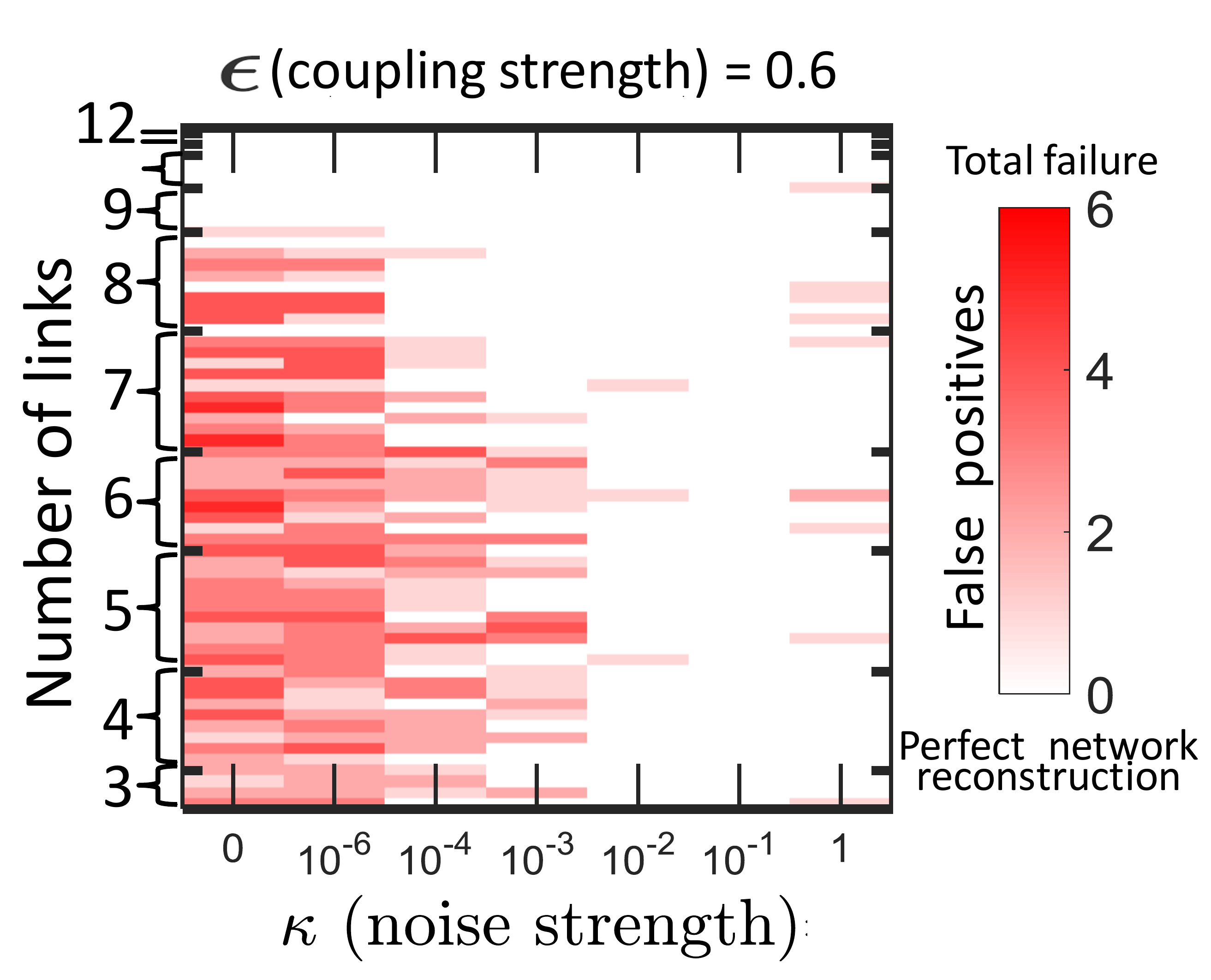}\\
{\bf(b)}\\
\includegraphics[scale=0.4,trim={0cm 0cm 0cm 0cm},clip]{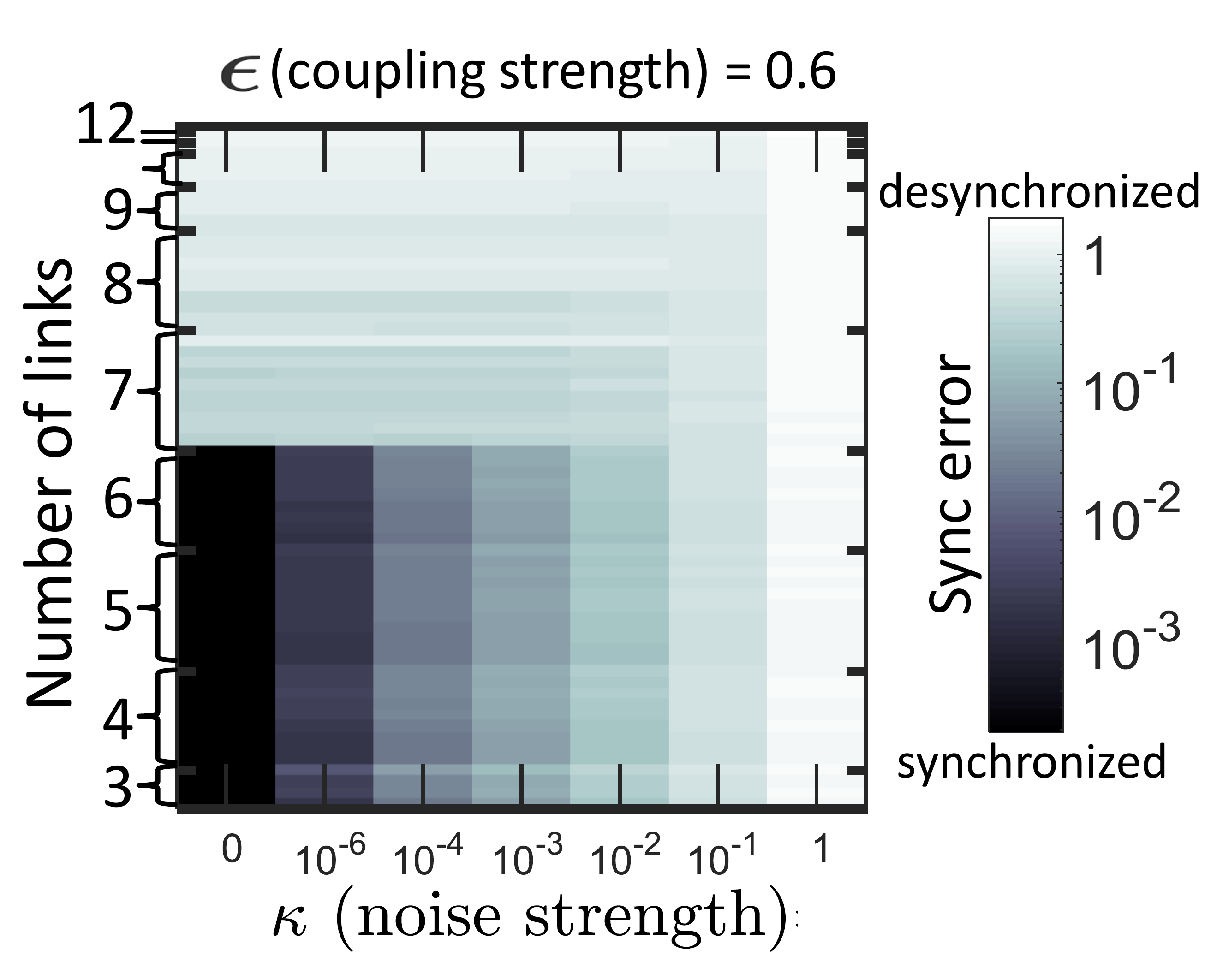}

\caption{Simulation test results with varying noise strength $\kappa$. (a) Number of false positives and (b) synchronization error for simulated time series from different networks with progressively increasing noise. As described in the text, each horizontal cut of the plots represents a single trajectory of the system, starting from a random initial condition. The convention for sequence of the networks is the same as in Fig. 5.}
\end{figure}

\subsection{Performance on Simulated Data - Homogeneous Delays}
In this section, we test our methodology on simulated time series for our coupled opto-electronic oscillator networks where links have identical delays. We will use these simulation tests to study the effects of noise and coupling strength on the amount of synchrony in the system, and their effect on the performance of link inference tasks. In particular, in Sec. 3.2 we showed that our opto-electronic oscillator networks show synchronized dynamics for certain ranges of the coupling strength $\epsilon$. As we will now show, our method works excellently when the system dynamics does not show pronounced global synchrony, while it does not work well when there is pronounced global synchrony. Furthermore, we will show that our technique gives excellent results when there is a small amount of dynamical noise present so that the global synchrony among the opto-electronic oscillators is appreciably broken. 

In order to directly demonstrate the effect of loss of synchrony on link inference, we vary the noise level and coupling strength in the following two sets of examples. In the first example set, the system starts with a random initial condition with no noise for $5\times 10^4$ time steps (about 1470 delay times) and is allowed to settle down to an attractor. Then we continue the simulation, but with the noise strength $\kappa$ set to $10^{-6}$ for the next $5\times 10^4$ steps, then with the noise strength set to $\kappa=10^{-4}$ for the next $5\times 10^4$ steps, and so on, keeping the coupling strength fixed at $\epsilon=0.6$ (Figs. 6(a) and 6(b)). As shown in Fig. 6(b), as the noise strength increases, it drives the system away from the attractor and disrupts its global synchrony, resulting in larger synchronization error, which allows better link inference performance, as shown in Fig. 6(a). In these figures, for each of the time series segments with a fixed noise strength, we use the first $3\times10^4$ time steps (about 880 delay times) to train our RC and infer links using our procedure described in Sec. 2. We repeat this process for each of the 62 possible connected networks of 4 opto-electronic oscillators \cite{ravoori2011robustness}, each one with a different random initial condition.

\begin{figure}[!ht]\label{increasing_epsilon}
{{\bf(a)}}\\
\includegraphics[scale=0.4,trim={0cm 0cm 0cm 0cm},clip]{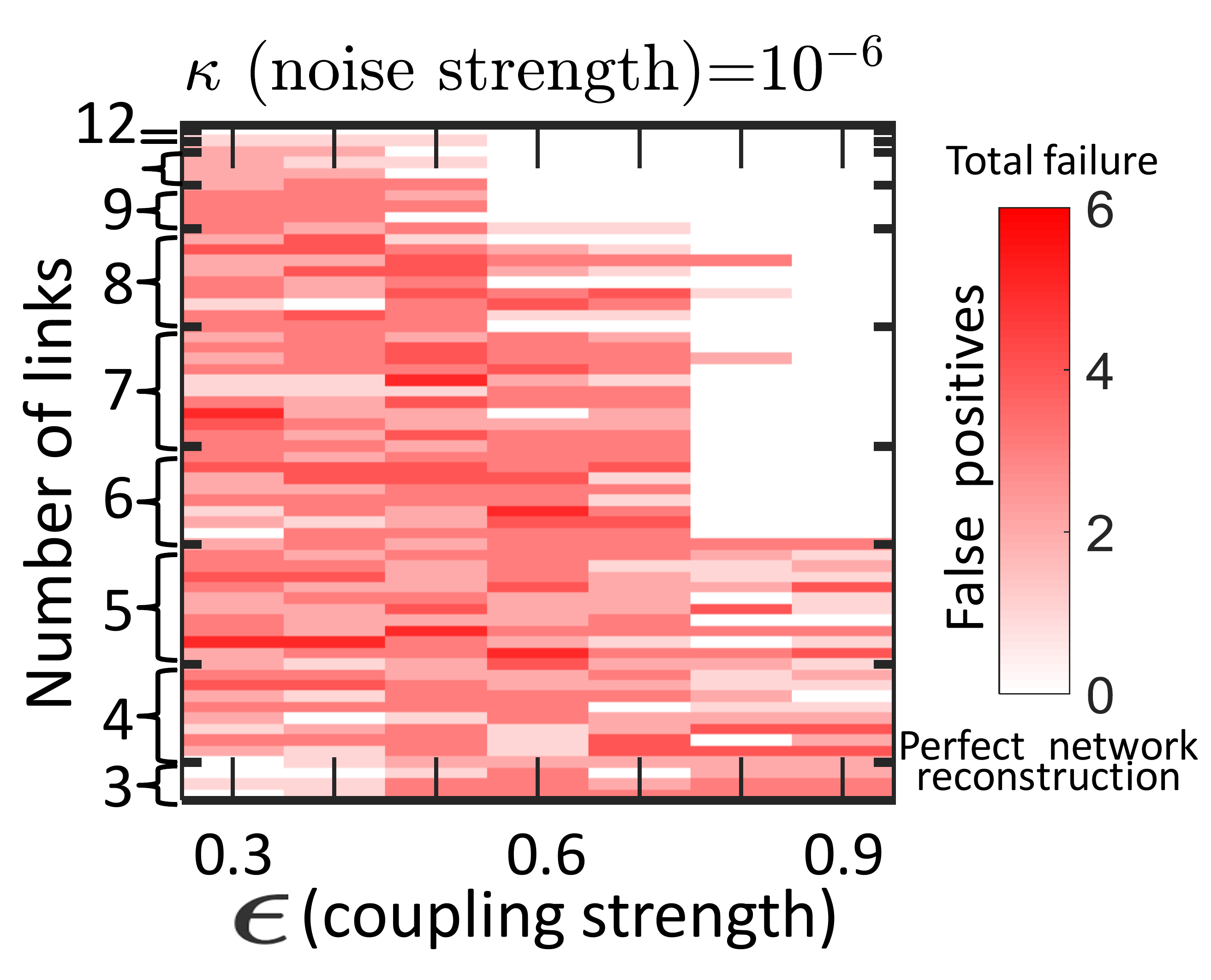}\\
{{\bf(b)}}\\
\includegraphics[scale=0.4,trim={0cm 0cm 0cm 0cm},clip]{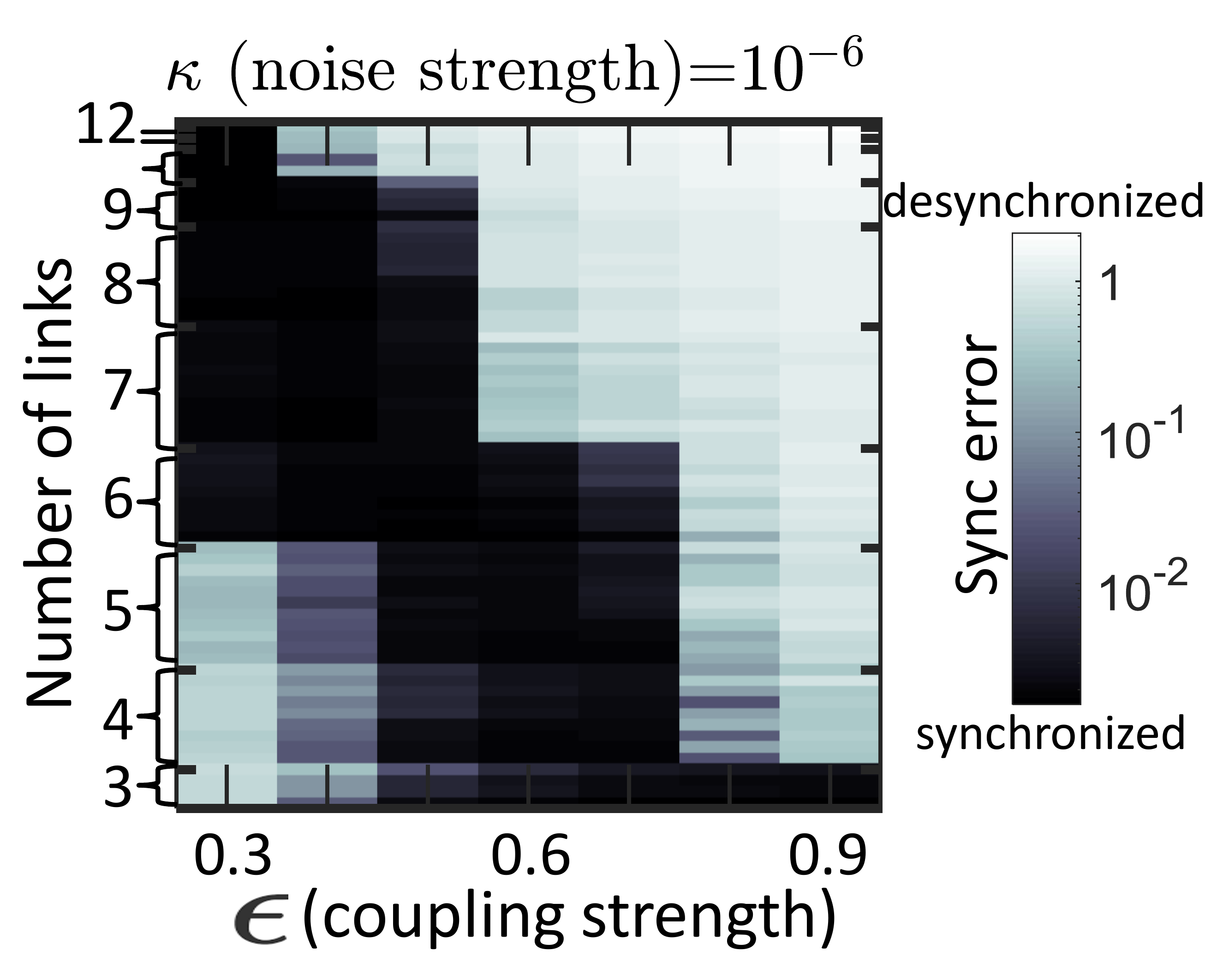}

\caption{Simulation test results with varying coupling strength $\epsilon$. (a) Number of false positives and (b) synchronization error for simulated time series from different networks with progressively increasing coupling strength. As described in the text, each horizontal cut of the plots represents a single trajectory of the system, starting from a random initial condition. The convention for sequence of the networks is the same as in Fig. 5.}
\end{figure}

We use the same procedure in the next set of examples (Figs. 7(a) and 7(b)), but this time we keep the noise level fixed at its nominal experimental value of $\kappa=10^{-6}$ and vary the coupling strength $\epsilon$ stepwise. The results are summarized in Fig. 7. For both Figs. 6 and 7, we simultaneously plot the number of false positives and synchronization error and follow the same convention as in Fig. 5.

As we see from Fig. 6, a greater degree of global synchronization generally corresponds to a larger number of false positives, consistent with the hypothesis that global synchrony is detrimental to the performance of link inference. This is expected because exact synchronization makes the time series from the four opto-electronic oscillators indistinguishable and hence the observed dynamics yields no information about their underlying causal interactions. In particular, we see that networks with eight or more links do not synchronize sufficiently even in absence of the noise, and we are indeed able to infer the links well, with, at most, only one false positive. In contrast, for networks with smaller numbers of links, which are strongly synchronized for noise levels $\kappa \lesssim 10^{-3}$, we have many false positive link inferences. Again, as we see from Fig. 6, all the networks show a loss of global synchrony for sufficiently strong noise levels ($\kappa \gtrsim 10^{-2}$) and this results in almost perfect link inference until the noise strength becomes significant compared to the noiseless opto-electronic oscillator signal amplitudes ($\kappa \gtrsim 10^0$). Other examples of similar beneficial role of noise in link inference can be found at earlier works as well, e.g., in \cite{banerjee2019using,panaggio2019model,leguia2019inferring,lipinski2015using,prill2015noise,ren2010noise,wang2012reverse}. {\color{black} Of these, Ref. \cite{wang2012reverse} describes a network inference technique based solely on noise correlations. However, Ref. \cite{wang2012reverse} proposes a technique applicable only in the case of Laplacian coupling, unlike our work which does not employ this model-specific restriction. Furthermore, unlike our method, their method does not work in the absence of dynamical noise, and was not validated using experimental data.}

In the second set of examples (Figs. 7(a) and 7(b)), we fix the noise at a particular strength $\kappa=10^{-6}$, and progressively increase the coupling strength $\epsilon$. We estimate that this noise level approximates that for the experimental tests reported in the next subsection (Sec. 4.2). As in the previous examples, Fig. 7 shows that our link inference method performs well when the global synchrony is not too strong. A difference in this set of results from the previous ones is the non-trivial relationship between the coupling strength $\epsilon$ and global synchrony, which we have already discussed in the last section (Fig. 5). Furthermore, we notice that, even in the absence of global synchrony, the coupling strength needs to exceed a minimum value (about 0.1) for successful link inference. For smaller coupling, the off-diagonal elements of the matrix ${\bf {M}}[n\Delta t]$ could be so small in magnitude that the values corresponding to actual links are of the same order as those corresponding to absent links. Thus, sufficiently large coupling strength $\epsilon$ is beneficial for our link inference technique because of the better contrast among the elements of ${\bf {M}}[n\Delta t]$ and diminished global synchrony.

%{\color{green} In the following paragraph, we need to be more precise abut what we are plotting. For what networks/parameters are these scores computed? Is it the scores used to obtain Fig. 6?}

\begin{figure}[!ht]\label{increasing_epsilon}
\includegraphics[scale=0.45,trim={0cm 0cm 0cm 0cm},clip]{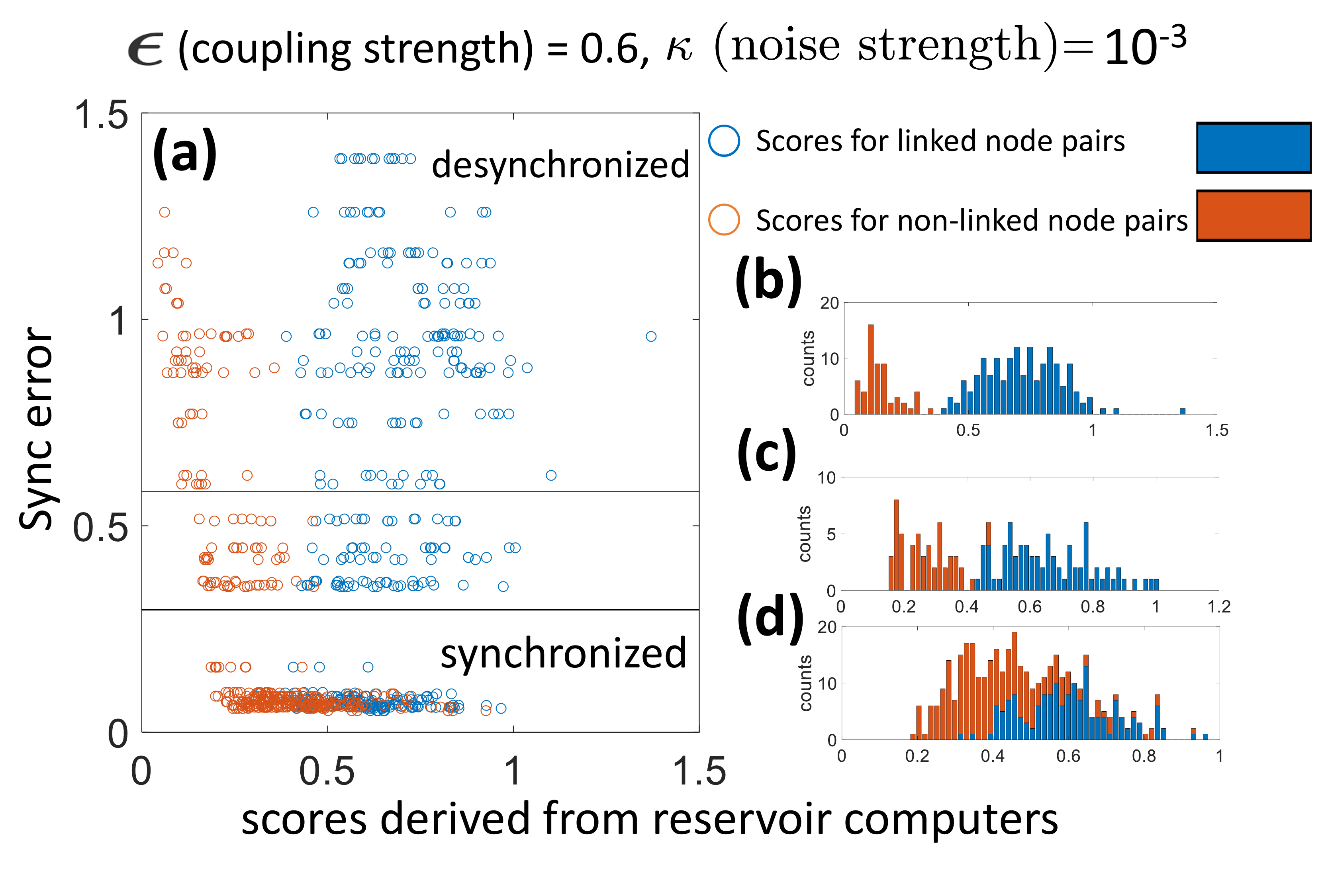}
\caption{{\color{black}This figure shows statistics of our results for the link  scores $S_{i,j}$ of all the possible networks (listed in Fig. 4) with $\epsilon=0.6$ and $\kappa=10^{-3}$, where, in determining the statistics, for each of the networks, we use a single random realization of the initial condition and reservoir couplings. In panel (a), for each individual directed node pair $(i,j)$, i.e., each candidate link, a point is plotted in the sync-error/score plane with true links plotted in blue, and link absences denoted in red, with red overlaying blue. The two black horizontal lines divide the sync-error/score plane into three regions corresponding to networks that are highly desynchronized, moderately synchronized, and strongly synchronized.
Panels (b), (c), and (d) show histograms of the node scores for each of the three levels of network synchronization demarcated in panel (a).
Bins with scores that all correspond to true links (absence of links) are colored blue (red).
Bins with scores corresponding to both true links and link absence are vertically stacked into upper and lower pieces where the lower piece (blue) corresponds to the number of true links in the bin, and the upper piece (red) corresponds to the number of missing links.} }
\end{figure}

{\color{black} {\it When the number of links is unknown.} Finally, to show the effectiveness of our procedure in situations where the number of links $L$ is unknown, in Fig. 8 we plot the distribution of the scores $S_{ij}$ calculated from the matrix ${\bf {M}}[n\Delta t]$, {\color{black} for all the $62$ networks listed in Fig. 4 with coupling strength $\epsilon=0.6$ and noise strength $\kappa=10^{-3}$, where, in determining the statistics, for each of the networks, we use a single random realization of initial condition and reservoir couplings}. Fig. 8 also shows the numerical values and properties of the scores we typically get in our method. In Fig. 8(a), we have labeled individual scores into two types [those corresponding to actual network links (colored blue), and those corresponding to absence of links (colored red)] and plotted the scores with the respective synchronization errors in the network. We see that, in the cases with complete desynchronization, where we obtain perfect network inference with known $L$ (as seen in Fig. 6(a)), it is also easy to predict a binary decision threshold on our scores (the top histogram, Fig. 8(b)). These scores separate into two distinct populations according to their magnitudes, with a gap in between them (Fig. 8(b)), and the populations correspond to actual links or absence of links. In the cases for which Fig. 6a shows finite but small number of false positives with known $L$, the two populations overlap, but there are still two distinct histogram peaks (the middle histogram, Fig. 8(c)), so that one can expect good inference results in cases where $L$ is not known by choosing a suitable threshold based on the shape of the histogram. However, when the histogram overlap is so great that two peaks are not discernible, we expect that the ability to infer links no longer exists (the bottom histogram, Fig. 8(d)). In Fig. 6a, this scenario corresponds to cases with a large number of false positives, even with a known $L$, because the network exhibits global synchrony.}

%{\color{green} I'm not sure what the last sentence is trying to say. Is it referring to Fig. 6 and Fig. 8?--JDH
%I got rid of that sentence -- AB}

\subsection{Performance on Experimental Data}

\begin{figure}[hbt!]\label{expt_histogram}
 \centering

\begin{centering}
\includegraphics[scale=0.32,trim={0cm 0cm 0cm 0cm},clip]{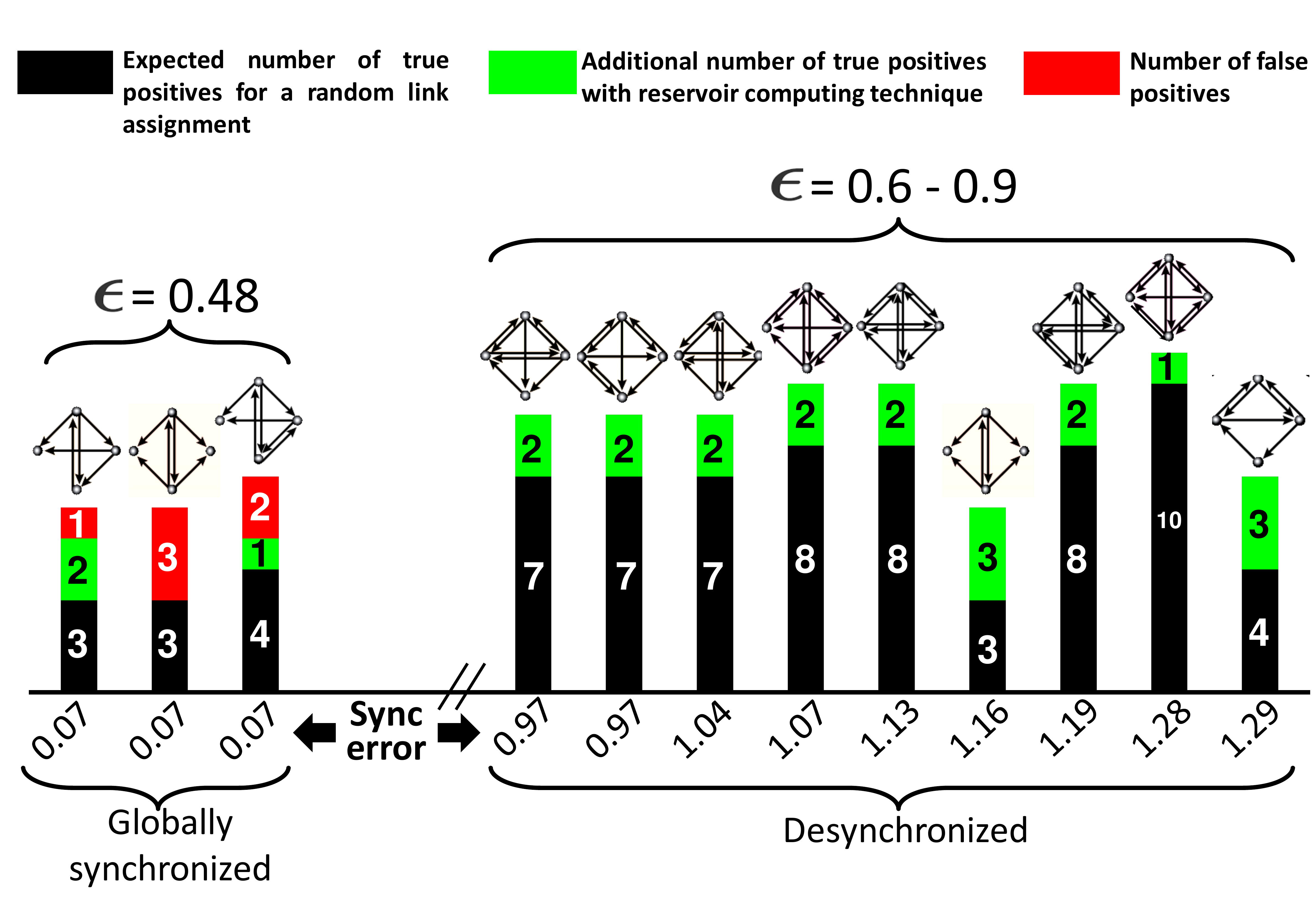}\caption{Typical performance of our network inference method on experimental data. Each of the vertical bars correspond to the experimental realization of a distinct opto-electronic oscillator network, with the height indicating the corresponding number of links. Each bar is separated into 3 segments as shown, but in many of them only two are seen if there is a perfect link inference. The numbers inside the segments indicate the corresponding heights of the segments, rounded to the nearest integers. Ranges of coupling strength ($\epsilon$) and synchronization error for all the examples are indicated as well}.
\end{centering}
\end{figure}

Having established the usefulness of our network inference method on simulated time series, we now report our experimental tests on the opto-electronic oscillator networks described in Sec. 4.1. In Fig. 9, we show some representative examples of the performance of our method on experimental time series. Each column in the figure corresponds to a time-series from a distinct network indicated above the column, with the respective global synchronization error indicated on the horizontal axis. The height of the columns gives the total number of links in the corresponding network. The columns are each divided into three parts (colored in red, green, and black in the plot). The height of the red portion indicates the number of falsely inferred links (``false positives", FP). This portion is absent in the many cases where we have perfect network inference. The number of correctly inferred links (``true positives", TP) is indicated by the total height of the green plus black portions of a column. The height of the black portion indicates the expected number of true positives on average (to nearest integer) that would be obtained if all $L$ links were to be guessed randomly, while the height of the green portion indicates the increase of true positives over what would result from random selection. To evaluate the expected number of randomly selected true positives, we note that, for $L$ links randomly and uniformly assigned among the $D_n\left ( D_n-1\right )$ ordered pairs of $D_n$ nodes, the expected number of false positive links is $L\left(1-\frac{L}{D_n\left ( D_n-1\right )}\right)$ and the expected number of true positive links is thus ${L^2}/{D_n\left ( D_n-1\right )}$. If our method yields more true positives than ${L^2}/{D_n\left ( D_n-1\right )}$, then we consider our method to be successful, even if it gives some false positives. 

To summarize our experimental results, consistent with the simulation results of Figs. 5 and 7, the time-series from the experimental opto-electronic oscillator networks (Fig. 9) were either globally well synchronized or else were strongly desynchronized, and, when strong desynchronization applied, our method correctly identified all of the links.

{
\color{black}
\subsection{Performance on Simulated Data - Heterogeneous Delays}

So far we have considered cases where the link time delays $\tau_{ij}^{c}$ in Eq.(1) are the same along all links  (i.e.,$\tau_{ij}^{c}=\tau$). We now present results on simulated data for which the link delays $\tau_{ij}^{c}$ are chosen randomly from a uniform distribution of mean $\tau_0$ and width $2\Delta \tau$, where the mean link delay $\tau_0$ is assumed known. We then apply our previously described method treating all links as if they had the same delay $\tau_0$, and assess the results as a function of the link delay heterogeneity, as characterized by the fractional spread $2\Delta \tau/\tau_0$ of the link delays. We do this for all networks listed in Fig. 4. For purposes of discussion, we divide these networks into two categories: (1) networks for which we obtained perfect inference with homogeneous delays, and (2) networks for which synchronization hindered link inference performance with homogeneous delays (Fig. 6a). For a fixed mean delay time $\tau_0$ (corresponding to $k=34$ with $\tau_0=k\Delta t$), the results for different amounts of spread of the delay times $\Delta \tau$ are plotted in Fig. 10. The results indicate that in case (1), we continue obtaining good results, with the average of maximum number of false positives around 1, if the heterogeneity of the spread in delays is not too large (Fig. 10a). In case (2), we obtain better, and, in some cases perfect (i.e., zero false positives), results with moderate delay heterogeneity. This improvement can be attributed to a change in network dynamics: The heterogeneity of the link delays inhibits global synchronization, as is evident from the corresponding synchronization error plot of Fig. 10b. Thus, moderate delay heterogeneity can be beneficial to the working of our method. This study confirms that our formalism can be applied to realistic networks with distributed link delay time, broadening the scope of our methodology. Further investigations in which we include the delay heterogeneity in the reservoir computer formalism itself are reserved for the future.
}

\begin{figure}[!ht]\label{increasing_epsilon}
{\bf(a)}\\
\includegraphics[scale=0.35,trim={0cm 0cm 0cm 0cm},clip]{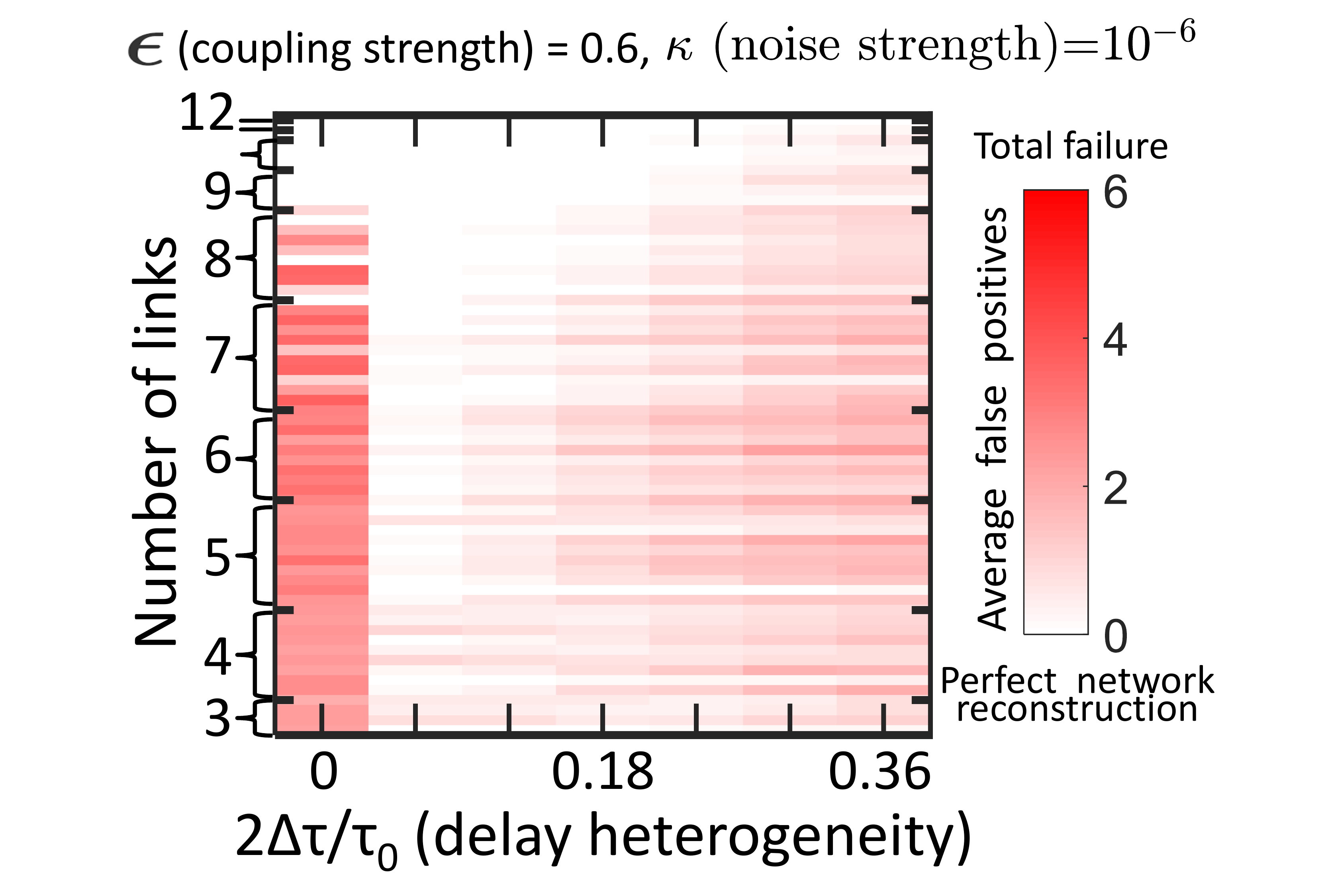}\\
{\bf(b)}\\
\includegraphics[scale=0.35,trim={0cm 0cm 0cm 0cm},clip]{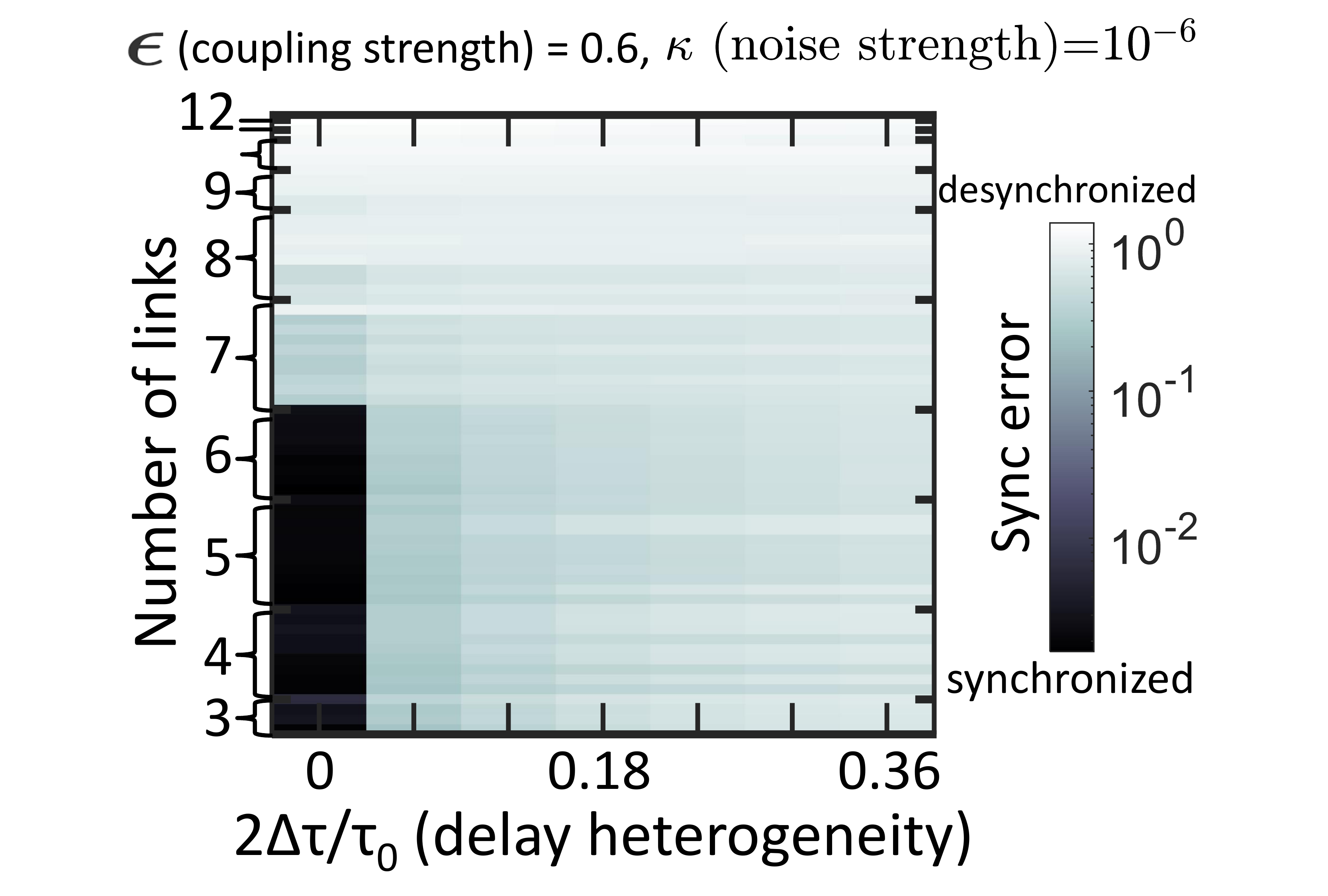}
\caption{{\color{black}Tests on networks with heterogeneous link delays. (a) Average number of false positives, and (b) the corresponding synchronization error for different networks listed in Fig. 4, with $\epsilon=0.6$ and $\kappa=10^{-6}$,  and different amounts of link delay heterogeneity, presented as the ratio of link delay variation range $2\Delta \tau$ and mean link delay $\tau_0$. The {\color{black}results are based on simulated data with different network configurations, and different random link delays along the network links.} In each case, the averaging is done over $100$ different random realizations of reservoir connections, initial conditions and assignments of interaction delays to different links. }}
\end{figure}

\section{Discussion}

In this work, we developed a reservoir-computer-based technique for the general problem of link inference of noisy delay-coupled networks from their nodal time series data and demonstrated the success of our method on simulated and experimental data from opto-electronic oscillator networks with identical and distributed link delay times. Our main findings are as follows:
\begin{itemize}

%\item Our network inference method is highly interdisciplinary in that it is inspired by the corresponding ideas from widely used invasive ``knock-down" techniques from genetics and molecular biology, but it implements these ideas in a non-invasive way, incorporating them on machine learned {\it in silico} models of unknown systems rather than the systems themselves. This approach retains the practicality of the direct invasive methods but greatly generalizes the scope of their applicability. 

\item Testing on experimental and simulated time-series datasets from networks, we found that, in the absence of dynamical noise, 
our method yields extremely good results, as long as there is no synchrony in the system.

\item We found that dynamical noise {\color{black} and/or a moderate amount of link time delay heterogeneity} can greatly enhance the performance of our method when synchrony is present provided that the noise amplitude {\color{black}or link time delay heterogeneity} is large enough to perturb the synchrony. 

\item Since dynamical noise is ubiquitous in natural and experimental situations, we anticipate that this technique may be useful in network inference tasks relevant to fields like biochemistry, neuroscience, ecology, and economics.
\end{itemize}

Among the important issues for future investigation, our work in this paper could be extended to cases when the dynamics of the network nodal states are partially synchronized (e.g., cluster synchronization of nodes \cite{pecora2014cluster}) or display generalized synchronization \cite{rulkov1995generalized,kocarev1996generalized,senthilkumar2008transition,senthilkumar2013global}. Effects of network symmetries \cite{whalen2015observability,pecora2014cluster}, non-uniform coupling \cite{denker2004breaking}, and non-identical nodes \cite{nishikawa2003heterogeneity} \textemdash all of which can affect the synchrony of nodal states \textemdash would also be very interesting to study. Another important issue that awaits study is the effects of incomplete \cite{han2015robust}, or erroneous nodal state data \cite{peixoto2018reconstructing, napoletani2008reconstructing} on link inference {\color{black}, e.g., a case of particular interest is that where measured nodal time series is only available from a subset of $N'<N$ of the $N$ network nodes, and the value of $N$ itself is unknown.} 

\section{Acknowledgements}
The authors would like to thank Thomas E. Murphy for his efforts to enable the experiments to be performed at the University of Maryland, College Park in the time of COVID-19. This work was supported by the U. S. National Science Foundation Grant DMS 1813027 (AB and EO) and the Office of Naval Research Grant N000142012139 (RR). AB acknowledges that he/they did this work at the Unversity of Maryland and its neighborhoods in College Park, which stand on the traditional, ancestral and contemporary lands of the Nacotchtank and Piscataway Peoples.

\section*{Appendices}
\renewcommand{\thesubsection}{\Alph{subsection}}
\renewcommand{\theequation}{\thesubsection.\arabic{equation}}
\setcounter{equation}{0}
\subsection{Determination of the time-delay from cross-correlation}

\begin{figure}[hbt!]\label{expt_histogram}
 \centering

\begin{centering}
\includegraphics[scale=0.4]{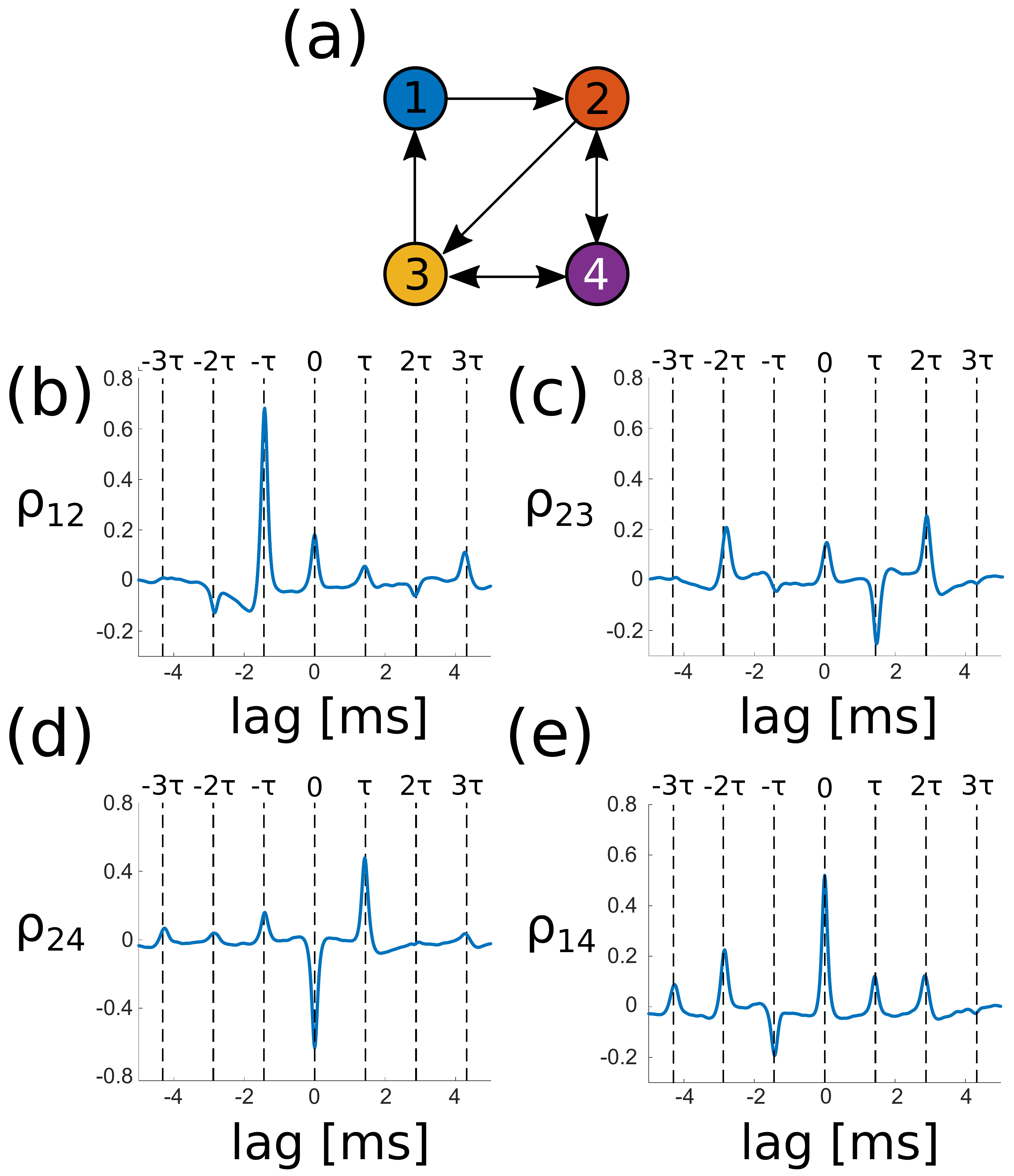}\caption{Time delay determination by cross-correlation of measured time series. In all cases, the coupling delay time is 1.44ms. (a) The network to be inferred from time series measurements. (b) The cross-correlation between the time series of nodes 1 and 2. A strong peak is observed near -1.44ms. (c) The cross-correlation between the time series of nodes 2 and 3. A strong negative peak is observed near 1.44ms and a strong positive peak near -1.44ms. (d) The cross-correlation between the time series of nodes 2 and 4. A strong peak is observed near 1.44ms. (e) The cross-correlation between the time series of nodes 1 and 4. Peaks are observed at 1.44ms and -1.44ms, even though there is no link from node 1 to node 4. \label{fig:xcorr}}
\end{centering}
\end{figure}

In this Appendix, we demonstrate that the duration of the delay along a link in our network can be accurately estimated from the cross-correlation between the measured time series of the two nodes connected by that link. In particular, we show that the location of the peak of the cross-correlation between the two nodes provides a good estimate of the delay time. We also show that the cross-correlation cannot determine causality, because it cannot determine the direction of a given putative link, or if the link even exists at all.

Consider the network depicted in Fig. \ref{fig:xcorr}a, where each node is an opto-electronic oscillator as described in Sec. 3.  The delay in each link is $\tau=1.44$ms. We define the cross-correlation between the sampled time series of two nodes $i$ and $j$ as
\begin{equation}
    \rho_{ij}(\mathrm{lag}) = \frac{1}{\sigma_i\sigma_j}\sum_k x_i[k+\mathrm{lag}]x_j[k],
\end{equation}
where $x_{i}[k]$ is the measured time series of node $i$ at discrete time $k$ and $\sigma_{i}$ is the RMS value of $x_i$. The time series should be mean-subtracted so that $\langle x_i\rangle=0$. The time series used here were obtained from experimental measurements of our opto-electronic oscillator network.

First, we compute $\rho_{12}$, the cross-correlation between node 1 and node 2, shown in Fig. \ref{fig:xcorr}b. A peak is located at -1.44ms. This corresponds to the delay in the link from node 1 to node 2. This suggests that the dynamics of node 2 lag behind the dynamics of node 1, as one might expect since the delayed link is from node 1 to node 2. 

Next, we compute $\rho_{23}$, the cross-correlation between node 2 and node 3, shown in Fig. \ref{fig:xcorr}c. The largest (negative, in this case) peak is located at 1.44ms, correctly identifying the absolute value of the delay time of the link from node 2 to node 3. However, in contrast to $\rho_{12}$, the peak location in $\rho_{23}$ shows that node 3 leads node 2. There is no peak at a lag of -1.44ms. This shows that the cross-correlation can identify the delay time, but not the link direction.

We now consider $\rho_{24}$. Nodes 2 and 4 have a bidirectional link; however, the cross-correlation $\rho_{24}$ shown in Fig. \ref{fig:xcorr}d has a prominent peak at 1.44ms but not at -1.44ms. There is no indication that the link is bidirectional.

Finally, we consider $\rho_{14}.$ There is no direct link between nodes 1 and 4. Still, the cross-correlation $\rho_{24}$ shown in Fig. \ref{fig:xcorr}e has peaks at both 1.44ms but at -1.44ms.

This example demonstrates that the cross-correlation can provide an accurate estimate of the duration of the delay in the coupling between two nodes, but that it does not provide sufficient information to determine the existence or directionality of a link. We find similar results in all the networks of opto-electronic oscillators we tested.

\subsection{Derivation of the discrete-time equation for simulating the opto-electronic system}
\setcounter{equation}{0}
In this appendix, we derive the discrete-time equations implemented by the DSP board in our experimental setup and used in our simulations. These discrete-time equations are derived using standard techniques for approximating an analog filter as a digital filter and are essentially a trapezoid rule approximation to Eqs. \ref{eq_u} and \ref{eq:u_mats}.

The derivation here closely follows that presented in Ref. \cite{murphy2010complex}. The missing details from Ref. \cite{murphy2010complex} are filled in here, drawing from Ref. \cite{oppenheim2001discrete} for the details of the z-transform and bilinear transform.

The continuous-time filter equations that describe a two-pole bandpass filter are

\begin{equation}\label{eq:filt_u}
    \frac{d\mathbf{u}(t)}{dt}=\mathbf{Eu}(t)-\mathbf{F}r(t)
\end{equation}
\begin{equation}
\label{eq:filt_x}
    x(t)=\mathbf{G}\mathbf{u}(t)
\end{equation}
where
\begin{equation}
\label{eq:x_mats}
    \mathbf{E}=\begin{bmatrix}-(\omega_L+\omega_H) & -\omega_L \\ \omega_L & 0\end{bmatrix}, \qquad
    \mathbf{F}=\begin{bmatrix}\omega_L\\0 \end{bmatrix}, \qquad\mathrm{and} \;\;\mathbf{G} = \begin{bmatrix} 1 & 0 \end{bmatrix}.
\end{equation}

Here, $\mathbf{u}(t)$ is a 2-vector that describes the state of the filter, $r(t)$ is the filter input, and $x(t)$ is the filter output. In the case of one of our opto-electronic oscillators $r(t) = \beta\cos^2(x(t-\tau) + \phi_0)$. In order to implement this filter digitally, one derives the digital filter equations by computing the transfer function of the analog filter, then applying the bilinear transform with frequency pre-warping to the continuous-time transfer function to obtain the discrete-time transfer function. From there, the discrete-time digital filter equations can be written down.

The transfer function $H(s)$ of the analog filter can be found by $H(s)\equiv X(s)/R(s)$, where the capital letters $X$ and $R$ indicate the Laplace transform of $x$ and $r$, respectively. We compute the Laplace transform of Eq. \ref{eq:filt_u}:

\begin{equation}
    s\mathbf{U}(s)=\mathbf{E}\mathbf{U}(s)+\mathbf{F}R(s) 
    %&\Rightarrow \mathbf{U}(s)=(s\mathbf{I}-\mathbf{E})^{-1}\mathbf{F}R(s)
    \label{eq:laplace}
\end{equation}

Then, performing the Laplace transform of Eq. \ref{eq:filt_x} and inserting Eq. \ref{eq:laplace}, we have:
%\begin{align}
%    X(s) = \mathbf{G}\mathbf{U}(s)=\mathbf{G}(s\mathbf{I}-\mathbf{E})^{-1}\mathbf{F}R(s).
%\end{align}
%We can rearrange this to obtain the filter transfer function:
\begin{equation}
    H(s)\equiv \frac{X(s)}{R(s)} = \mathbf{G}(s\mathbf{I}-\mathbf{E})^{-1}\mathbf{F}
%%\end{equation}
% the matrices from Eq. \ref{eq:x_mats}, we have
%\begin{equation}
    =\frac{s\tau_H}{(1+\tau_Ls)(1+\tau_Hs)},
    \label{eq:analog_transfer}
\end{equation}
where $\tau_H=1/\omega_H$ and $\tau_L=1/\omega_L$. Equation \ref{eq:analog_transfer} is the continuous time filter transfer function for the filter described by Eqs. 
\ref{eq:filt_u}-\ref{eq:x_mats}.

Two standard tools used in the design and analysis of digital filters are the z-transform and the bilinear transform. The z-transform is the discrete time analog of the Laplace transform. The bilinear transform is a tool used to turn a continuous-time filter into a discrete-time filter. It can be shown that the result obtained by the bilinear transform method we use here is equivalent to applying the trapezoidal integration rule to Eqs. \ref{eq:filt_u}-\ref{eq:x_mats} \cite{oppenheim2001discrete}.

The z-transform is defined as
\begin{equation}
    Z\{x[n]\}\equiv\sum_{n=-\infty}^{\infty}x[n]z^{-n},
\end{equation}
where $z$ is a continuous complex variable, and $n$ is discrete time. One important z-transform relations is that a delay by $m$ time steps in the discrete-time domain is equivalent to multiplication by $z^{-m}$ in the z-domain.  %One trivial, but important, z-transform relation is 
%\begin{equation}
%\label{eq:z-identity}
%    Z\{\delta[n-m]\}\equiv\sum_{n=-\infty}^{\infty}\delta[n-m]z^{-n}=z^{-m}.
%\end{equation}
%In other words, 

The bilinear transform is used to convert our continuous-time filter transfer function (Eq. \ref{eq:analog_transfer}) into a discrete-time filter transfer function. An exact conversion is done by discretizing with a time-step of $T$ and equating $z=e^{sT}$. Since $T$ is small, we can approximate

\begin{equation}
\label{eq:bilinear}
    s=\frac{1}{T}\ln(z)=\frac{2}{T}\frac{1-z^{-1}}{1+z^{-1}}.
\end{equation}
Equation \ref{eq:bilinear} is the bilinear transform.  This approximation is equivalent to applying the trapezoid rule to the continuous-time filter equations \cite{oppenheim2001discrete}. When Eq. \ref{eq:bilinear} is substituted into Eq. \ref{eq:analog_transfer}, we obtain the transfer function for a discrete-time filter with similar characteristics to the desired analog filter:

\begin{equation}
    H(z)=\frac{1}{4}(1-z_L)(1+z_H)\frac{1-z^{-2}}{(1-z_Lz^{-1})(1-z_Hz^{-1})}.
\end{equation}

This change of variables is a nonlinear mapping, so frequency warping occurs. This effect is minimal when the frequencies are significantly less than the Nyquist frequency (in this case $f_L=2.5$kHz and the Nyquist frequency is 12kHz) and can be further mitigated by pre-warping the frequencies of the continuous-time filter by $\Omega=\frac{2}{T}\tan(\frac{\omega}{2})$, where $\Omega$ is the discrete-time frequency and $\omega$ is the continuous-time frequency \cite{oppenheim2001discrete}. Therefore, we find that

\begin{equation*}
     z_H=\frac{1-\tan(T/2\tau_H)}{1+\tan(T/2\tau_H)}
\end{equation*}
and
\begin{equation*}
     z_L=\frac{1-\tan(T/2\tau_L)}{1+\tan(T/2\tau_L)}.
\end{equation*}

Now, one can use the definition of the transfer function $H(z)\equiv X(z)/R(z)$ to find
\begin{equation}
\label{eq:intermediateZ}
    (1-(z_L+z_H)z^{-1}+z_Lz_Hz^{-2})X(z)=\frac{1}{4}(1-z_L)(1+z_H)(1-z^{-2})R(z).
\end{equation}

We arrive at the discrete-time filter equation by performing the inverse z-transform on Eq. \ref{eq:intermediateZ}:% (taking advantage of Eq. \ref{eq:z-identity}):

\begin{equation}
    x[n]=(z_L+z_H)x[n-1]-z_Lz_Hx[n-2]+\frac{1}{4}(1-z_L)(1+z_H)\left(r[n]-r[n-2]\right).
\end{equation}
For the filter used in this work, $z_L+z_H$=1.4845, $z_Lz_H$=0.4968, and $\frac{1}{4}(1-z_L)(1+z_H)$=0.242.

\printbibliography

\end{document}